\PassOptionsToPackage{unicode}{hyperref}
\PassOptionsToPackage{hyphens}{url}
\PassOptionsToPackage{dvipsnames,svgnames,x11names}{xcolor}
\documentclass[
  a4paper,
]{article}
\usepackage{xcolor}
\usepackage{amsmath,amssymb}
\usepackage{iftex}
\ifPDFTeX
  \usepackage[T1]{fontenc}
  \usepackage[utf8]{inputenc}
  \usepackage{textcomp} % provide euro and other symbols
\else % if luatex or xetex
  \usepackage{unicode-math} % this also loads fontspec
  \defaultfontfeatures{Scale=MatchLowercase}
  \defaultfontfeatures[\rmfamily]{Ligatures=TeX,Scale=1}
\fi
\usepackage{lmodern}
\ifPDFTeX\else
\fi
\IfFileExists{upquote.sty}{\usepackage{upquote}}{}
\IfFileExists{microtype.sty}{% use microtype if available
  \usepackage[]{microtype}
  \UseMicrotypeSet[protrusion]{basicmath} % disable protrusion for tt fonts
}{}
\usepackage{setspace}
\makeatletter
\@ifundefined{KOMAClassName}{% if non-KOMA class
  \IfFileExists{parskip.sty}{%
    \usepackage{parskip}
  }{% else
    \setlength{\parindent}{0pt}
    \setlength{\parskip}{6pt plus 2pt minus 1pt}}
}{% if KOMA class
  \KOMAoptions{parskip=half}}
\makeatother
\makeatletter
\ifx\paragraph\undefined\else
  \let\oldparagraph\paragraph
  \renewcommand{\paragraph}{
    \@ifstar
      \xxxParagraphStar
      \xxxParagraphNoStar
  }
  \newcommand{\xxxParagraphStar}[1]{\oldparagraph*{#1}\mbox{}}
  \newcommand{\xxxParagraphNoStar}[1]{\oldparagraph{#1}\mbox{}}
\fi
\ifx\subparagraph\undefined\else
  \let\oldsubparagraph\subparagraph
  \renewcommand{\subparagraph}{
    \@ifstar
      \xxxSubParagraphStar
      \xxxSubParagraphNoStar
  }
  \newcommand{\xxxSubParagraphStar}[1]{\oldsubparagraph*{#1}\mbox{}}
  \newcommand{\xxxSubParagraphNoStar}[1]{\oldsubparagraph{#1}\mbox{}}
\fi
\makeatother

\usepackage{longtable,booktabs,array}
\usepackage{calc} % for calculating minipage widths
\usepackage{etoolbox}
\makeatletter
\patchcmd\longtable{\par}{\if@noskipsec\mbox{}\fi\par}{}{}
\makeatother
\IfFileExists{footnotehyper.sty}{\usepackage{footnotehyper}}{\usepackage{footnote}}
\makesavenoteenv{longtable}
\usepackage{graphicx}
\makeatletter
\newsavebox\pandoc@box
\newcommand*\pandocbounded[1]{% scales image to fit in text height/width
  \sbox\pandoc@box{#1}%
  \Gscale@div\@tempa{\textheight}{\dimexpr\ht\pandoc@box+\dp\pandoc@box\relax}%
  \Gscale@div\@tempb{\linewidth}{\wd\pandoc@box}%
  \ifdim\@tempb\p@<\@tempa\p@\let\@tempa\@tempb\fi% select the smaller of both
  \ifdim\@tempa\p@<\p@\scalebox{\@tempa}{\usebox\pandoc@box}%
  \else\usebox{\pandoc@box}%
  \fi%
}
\def\fps@figure{htbp}
\makeatother

\NewDocumentCommand\citeproctext{}{}

\makeatletter
 \let\@cite@ofmt\@firstofone
 \def\@biblabel#1{}
 \def\@cite#1#2{{#1\if@tempswa , #2\fi}}
\makeatother
\newlength{\cslhangindent}
\newlength{\csllabelwidth}
\newenvironment{CSLReferences}[2] % #1 hanging-indent, #2 entry-spacing
 {\begin{list}{}{%
  \setlength{\itemindent}{0pt}
  \setlength{\leftmargin}{0pt}
  \setlength{\parsep}{0pt}
  \ifodd #1
   \setlength{\leftmargin}{\cslhangindent}
   \setlength{\itemindent}{-1\cslhangindent}
  \fi
  \setlength{\itemsep}{#2\baselineskip}}}
 {\end{list}}
\usepackage{calc}

\usepackage[margin=1in]{geometry}
\usepackage{setspace}
\usepackage{xcolor}
\usepackage{graphicx}
\usepackage{float}
\usepackage{tcolorbox}
\usepackage{wrapfig}
\usepackage{booktabs}
\usepackage{multirow}
\usepackage{multicol}
\usepackage{makecell}
\usepackage{siunitx}
\usepackage{tabularx}

\usepackage{amsmath, amssymb, amsfonts}
\usepackage{hyperref}
\usepackage[backend=biber]{biblatex}
\usepackage{lineno}
\usepackage{fancyhdr}
\fancypagestyle{plain}{
  \fancyhf{}
  \fancyhead[L,R]{\small\color{red}PREPRINT/WORKING PAPER}
  \fancyhead[C]{%
    \small\color{red}\url{https://krishpn.github.io}
  }
  \fancyfoot[L,R]{\small\color{red}PREPRINT/WORKING PAPER}
  \fancyfoot[C]{\thepage}
}

\makeatletter
\@ifpackageloaded{caption}{}{\usepackage{caption}}
\AtBeginDocument{%
\ifdefined\contentsname
  \renewcommand*\contentsname{Table of contents}
\else
  \newcommand\contentsname{Table of contents}
\fi
\ifdefined\listfigurename
  \renewcommand*\listfigurename{List of Figures}
\else
  \newcommand\listfigurename{List of Figures}
\fi
\ifdefined\listtablename
  \renewcommand*\listtablename{List of Tables}
\else
  \newcommand\listtablename{List of Tables}
\fi
\ifdefined\figurename
  \renewcommand*\figurename{Figure}
\else
  \newcommand\figurename{Figure}
\fi
\ifdefined\tablename
  \renewcommand*\tablename{Table}
\else
  \newcommand\tablename{Table}
\fi
}
\@ifpackageloaded{float}{}{\usepackage{float}}
\floatstyle{ruled}
\@ifundefined{c@chapter}{\newfloat{codelisting}{h}{lop}}{\newfloat{codelisting}{h}{lop}[chapter]}
\floatname{codelisting}{Listing}

\makeatother
\makeatletter
\@ifpackageloaded{caption}{}{\usepackage{caption}}
\@ifpackageloaded{subcaption}{}{\usepackage{subcaption}}
\makeatother
\usepackage{bookmark}
\IfFileExists{xurl.sty}{\usepackage{xurl}}{} % add URL line breaks if available
\hypersetup{
  pdftitle={The Information Dynamics of Insider Intent: How Reporting Inversions (Form 144) Mask Informational Rents in Insider Sales (Form 4)},
  pdfauthor={Krishna Neupane     ORCID: 0000-0003-1689-0557},
  colorlinks=true,
  linkcolor={blue},
  filecolor={Maroon},
  citecolor={Blue},
  urlcolor={Blue},
  pdfcreator={LaTeX via pandoc}}

\title{The Information Dynamics of Insider Intent: How Reporting
Inversions (Form 144) Mask Informational Rents in Insider Sales (Form
4)}
\author{Krishna Neupane \newline \vspace{0.5em}
\small \href{mailto:kneupan@gmu.edu}{kneupan@gmu.edu} \vspace{0.5em}
\small ORCID: 0000-0003-1689-0557}
\date{2026-02-19}
\begin{document}
\maketitle
\begin{abstract}
This study identifies and quantifies a significant informational
friction embedded in the SEC Form 144 disclosure regime, characterized
as predictive decoupling. Drawing on a theoretical foundation of welfare
economics, the article argues that the current reporting inversion
---where trade execution (Form 4) frequently precedes the public notice
of intent (Form 144)---violates the conditions for Pareto efficiency by
inducing non-symmetric pricing. Utilizing an event-study framework of
intent-to-sell windows, the analysis examines cases where insiders file
a notice of proposed sale but fail to execute within the statutory
90-day period. The machine learning audit reveals a persistent 52.4\%
opacity rate, where aborted signals remain statistically
indistinguishable from routine executions, creating a structural
information ceiling that prevents the market from exhausting the
signal's informational content. Contrary to the traditional small-firm
effect, the study documents a large-cap significance paradox- while
small-cap portfolios yield higher absolute abnormal returns (32.21 bps),
statistically significant alpha is concentrated in large-cap firms
(14.49 bps, \(p = 0.021\)). The results suggest that Institutional
Salience enables more reliable processing of this negative non-event
when reputational costs are maximized. Cross-sectional tests confirm
that prior idiosyncratic volatility serves as a signal amplifier, with
causal estimators identifying an illiquidity jump of up to 2.63 times.
To mitigate this market failure, the study proposes a mandatory
execution confirmation (Form 144-A) to transition the regime toward
bilateral accountability, converting a predictive blind spot into a
verifiable data stream and restoring the informational symmetry
requisite for efficient capital allocation.

\vspace{1em}

\noindent © Krishna Neupane 2026. All rights reserved. This working
paper is part of on going review for publication. No part of this
publication may be reproduced without prior permission.
\url{https://krishpn.github.io}

\vspace{1em}


\vspace{1em}

\end{abstract}

\setstretch{1.5}
\section{Introduction}

The First Fundamental Theorem of Welfare Economics (FTWE) posits that
competitive markets achieve Pareto efficient allocations under the
condition of symmetric information. Insider trading intent, generated
privately and subsequently disclosed via SEC Form 144 (Notice of
Proposed Sale of Securities), represents a fundamental violation of this
condition. While the Form 144 mandate is intended to restore market
symmetry, the temporal lag and execution uncertainty surrounding
proposed sales prevent the immediate satisfaction of FTWE conditions.
The resulting persistent information asymmetry induces a market failure
characterized by non-Pareto efficient pricing. Consequently, equity
prices during the filing window fail to reflect fundamental value,
representing an inefficient allocation of capital.

This theoretical breakdown is operationalized by the current regulatory
environment. As documented by Franzen, Li, and Vargus (2013), the
post-Sarbanes-Oxley (SOX) era is characterized by a reporting inversion;
trade execution (Form 4) is mandated within two business days, whereas
the notice of intent (Form 144) is frequently delayed. Historically,
Franzen, Li, and Vargus (2013) found that while Form 144 preceded Form 4
in 93.7\% of pre-SOX cases, this frequency declined to 17.5\% post-SOX.
This inversion serves as the empirical mechanism for market failure; at
the moment of execution, the prevailing market price cannot incorporate
the restricted nature of the securities being sold, resulting in a
significant transparency deficit.

Insiders may exploit this regulatory friction through strategic
disclosure sequencing. Although post-SOX regimes sought to
disincentivize reporting delays, insiders utilize the Filing Gap to
mitigate market impact. Prior research suggests that the stringency of
disclosure regimes directly influences opportunistic managerial trading
(Cheng, Nagar, and Rajan (2007), Huddart, Hughes, and Levine (2001),
Grossman (1981), Grossman and Stiglitz (1980)). Furthermore, heightened
scrutiny in the post-SOX environment may increase perceived legal
jeopardy, theoretically constraining the profitability of informative
trades.

This article argues that the reporting inversion facilitates a
sophisticated circumvention of these constraints, manifested as market
sentiment bias. Under the contrarian framework of Lakonishok and Lee
(2001) and the decoding logic of Cohen, Malloy, and Pomorski (2012),
insiders acting as behavioral contrarians typically execute sales during
periods of price appreciation. Because Form 144 is frequently reported
ex-post (Franzen, Li, and Vargus (2013)), insiders capitalize on signal
misclassification: the market may interpret a high-signal opportunistic
trade as a routine, liquidity-driven exit. By integrating with
uninformative trading volume, the insider preserves trade profitability
while minimizing the visibility that triggers regulatory or market
scrutiny.

The existence of this gap suggests delayed price discovery. As Kothari,
So, and Verdi (2016) note, if the initial market reaction to a signal is
biased or insufficient, subsequent price drifts or reversals become
predictable. By formalizing the 90-day non-execution event---defined as
the expiration of a Form 144 notice without a corresponding trade---this
analysis shifts the focus from realized transactions to omitted trades.
In this context, the market's initial failure to interpret the intent
ensures that informational content is not exhausted upon filing.
Instead, it manifests as a predictable negative drift (\(H_1\)) or a
subsequent reversal upon the trade's abandonment (\(H_2\)).

This study identifies a persistent non-execution frequency of 52.4\%,
where unexecuted notices remain statistically indistinguishable from
realized trades, establishing a structural barrier to market efficiency.
This allows for the quantification of informational rents: a window of
market inefficiency where insiders operate during periods of delayed
price correction. This concept extends the work of Zhang (2006), who
demonstrates that information uncertainty---proxied by fundamental
volatility---amplifies investor underreaction. Consistent with the
prediction that behavioral biases are most acute when assets are
difficult to value, empirical results show that prior idiosyncratic
volatility significantly amplifies the abnormal returns of unexecuted
signals (\(p = 0.021\)).

Finally, the informational potency of these signals challenges the
established small-firm effect. While traditional literature (Lakonishok
and Lee (2001), Seyhun (1986)) suggests insider signals are only
informative in small-cap stocks, this research identifies a
statistically significant 14.49 bps abnormal return (\(p = 0.021\)) in
large-cap sub-samples. This suggests that the 90-day non-execution event
possesses higher signal reliability in transparent, professionalized
environments. Market efficiency in large-cap equities does not preclude
the existence of abnormal returns; rather, it shifts the signal's nature
from a profit-seeking event to a high-stakes reputational one, where the
market eventually corrects the information contained within the
unexecuted intent.

To account for complex, non-linear dynamics, this manuscript implements
a multi-stage empirical framework. The analytical progression initiates
with baseline univariate OLS correlations, advances to high-dimensional
predictive modeling via Gradient Boosted Decision Trees (XGBoost), and
culminates in a Causal Machine Learning (CML) ensemble. This
methodological sequence is designed to isolate the causal mechanisms
underlying market salience, ensuring that the documented effects extend
beyond simple predictive correlation.

The remainder of the article is organized as follows. Section
\ref{sec:Central_Thesis_Contribution_Sample_Design} delineates the
central hypotheses and theoretical contributions. Section
\ref{sec:methodology} describes the three-phase methodological strategy:
establishing statistical baselines, evaluating the predictive
determinants of market liquidity, and identifying causal factors through
heterogeneous treatment effects. Section
\ref{sec:data_sample_construction} details the data construction
process, which integrates insider trading records, executive profiles,
and market-level financial metrics. Section
\ref{sec:results_and_analysis} presents the empirical findings and
subsequent analysis. Finally, Section
\ref{sec:conclusion_and_implications} offers concluding remarks.

\section{Hypothesis Development} \label{sec:Central_Thesis_Contribution_Sample_Design}

To empirically evaluate the welfare implications of information
asymmetry, this article utilizes the structural reporting windows
established by SEC Form 144 filings. Within the framework of the FTWE,
these filings serve as proxies for non-public signals that temporarily
impede the conditions requisite for a Pareto efficient competitive
equilibrium. To isolate these informational effects, the research design
partitions the dataset into two cohorts based on the subsequent filing
outcome:

\begin{description}

    \item[Hypothesis 1 ($H_1$):] The interval between the filing of Form 144 and the subsequent execution reported in Form 4 represents a period of incomplete price discovery. During this window, securities are expected to exhibit a persistent negative price drift as the market incorporates the information contained in the proposed divestment. This drift constitutes a measurable welfare cost resulting from information-induced price distortion.
    
    \item[Hypothesis 2 ($H_2$):] The expiration of a Form 144 window without a trade execution represents the withdrawal of the initial selling signal. This non-execution event triggers a price reversal as market participants adjust to the absence of the expected divestment. This correction facilitates a return to informational equilibrium by removing the price distortion established during the filing window.
\end{description}

Table \ref{tab:sample_partitioning} summarizes the data partitioning and
the expected signal transitions used to test the central hypotheses:

\begin{table}[h]
\centering
\caption{Framework for Testing Hypotheses: Mapping Filing Outcomes to Predicted Price Behavior}
\label{tab:sample_partitioning}
    \begin{tabular}{p{0.20\linewidth} | p{0.55\linewidth} | p{0.10\linewidth}}
        \toprule\textbf{Data Group} & \textbf{Economic Role (Signal Logic)} & \textbf{Hypothesis} \\
        \midrule\textbf{Executed Sales} & \textbf{Filing adds a signal $\rightarrow$ Price Drift}: Measures price distortion and the delay in market clearing during the active filing window. & $H_1$ \\
        \midrule\textbf{Unexecuted Notices} & \textbf{Expiration removes signal $\rightarrow$ Price Reversal}: Serves as a counterfactual to observe price correction after the selling intent is withdrawn. & $H_2$ \\
        \bottomrule
    \end{tabular}
\end{table}

\section{Methodology}\label{sec:methodology}

The empirical strategy is organized into a three-stage hierarchy
designed to move from descriptive association to causal identification.
This structure ensures that the documented market failures are not
merely predictive correlations but structural features of the regulatory
environment. The analysis initiates with univariate OLS and
high-dimensional fixed-effects models to establish the baseline
relationship between reporting intervals and abnormal returns. This
stage verifies the existence of the H\(_1\) negative drift and the
H\(_2\) price reversal, providing the initial statistical signal of
market mispricing during the Form 144 filing window.

To determine if the observed pricing inefficiency can be resolved by
market participants, Phase II implements a cross-paradigm audit. This
stage stress-tests the decodability of aborted intents across ten
distinct architectures, ranging from linear Elastic Net to
state-of-the-art FT-Transformers and Neuro-Symbolic TabNet. Rather than
identifying feature importance, this phase seeks to identify the
information ceiling---the point at which increasing model complexity
fails to improve predictive recall---thereby proving that the 52.4\%
Opacity Rate is a structural property of the public data. The final
stage employs a causal machine learning ensemble, utilizing double
machine learning and generalized random forest. This framework isolates
the Conditional Average Treatment Effect of an aborted trade. By
satisfying the unconfoundedness assumption through high-dimensional
controls and honest inference, Phase III confirms that the 1.3\% pricing
differential is a causal result of the abrupt disconnect, establishing a
formal link between the reporting inversion and Pareto inefficiency.

\subsection{Phase I: Baseline Statistical Significance and Univariate Analysis}\label{sec:phase_I_baseline_statistical_significance_univariate_analysis}

The analysis begins with a univariate test to evaluate market activity
during the reporting interval---the temporal window of unresolved
information between the notice of intent (Form 144) and the trade
execution (Form 4). This is quantified as the integer duration in
Equation \ref{eq:filing_gap_duration}:

\begin{equation}
    \label{eq:filing_gap_duration}
    G_{\text{days}, i} = \text{Date}_{\text{Execution}, i} - \text{Date}_{\text{Filing}, i},
\end{equation}

where \(G_{\text{days}, i}\) represents the duration of the
informational (filing) gap for filing \(i\), while
\(\text{Date}_{\text{Execution}, i}\) and
\(\text{Date}_{\text{Filing}, i}\) denote the calendar dates of the
trade execution and the initial notice of intent, respectively.

This duration serves as the active window for calculating the Cumulative
Market Excess Return (CMER). To capture the total wealth effect
occurring while the insider's intent remains unreflected in market
prices, the methodology employs geometric compounding. The CMER is
derived by aggregating daily market excess returns (\(R_{m,t}\)) from
the day following the filing (\(F+1\)) through the date of execution
(\(E\)) (Equation \ref{eq:cmer_calculation}):

\begin{equation}
    \label{eq:cmer_calculation}
    CMER_i = \left[ \prod_{t=F+1}^{E} (1 + R_{m,t}) \right] - 1,
\end{equation} where CMER\(_i\) is the compounded excess return for the
\(i\)-th filing, and \(R_{m,t}\) represents the market return in excess
of the risk-free rate on day \(t\). By defining the window as
\([F+1, E]\), the research ensures that the CMER reflects market
appreciation occurring after the intent is filed but before the
transaction is finalized. This approach utilizes geometric compounding
to avoid the upward bias inherent in the arithmetic summation of daily
returns, ensuring that the CMER accurately reflects the cumulative
wealth effect during the reporting interval (Lakonishok and Lee (2001)).
A one-sample \(t\)-test is performed against the null hypothesis
\(H_0: \mu_{CMER} = 0\) to confirm the existence of significant market
activity during the transaction cycle.

A statistically significant positive CMER suggests that insiders
strategically time executions to coincide with periods of market
appreciation. This behavior potentially masks opportunistic intent
(Cohen, Malloy, and Pomorski (2012)), as the idiosyncratic signal of the
trade is subsumed by the broader market performance during the reporting
interval. This behavior is further evaluated via the Divestment
Intensity (SM\(_i\)), which is defined as the ratio of shares sold to
total shares held (Equation \ref{eq:signal_magnitude}):

\begin{equation}
    \label{eq:signal_magnitude}
    SM_{i} = \frac{Q_{\text{sold}, i}}{Q_{\text{held}, i}},
\end{equation}

where \(Q_{\text{sold}, i}\) represents the number of shares sold in
transaction \(i\) (adjusted for stock splits), and
\(Q_{\text{held}, i}\) represents the total number of shares held by the
insider prior to the transaction. SM\(_i\) distinguishes between routine
liquidity rebalancing and high-conviction divestments. A high SM\(_i\)
coupled with a positive CMER identifies trades where informational
opacity is most critical to preserving the insider's execution price.

\subsection{Carhart Four-Factor Event Study}\label{sec:carhart_four_factor_event_study}

While the univariate CMER establishes the presence of price drift during
the reporting interval, it does not isolate the idiosyncratic alpha
generated by the insider's private information. To address this, a
firm-specific Carhart Four-Factor Event Study is employed to isolate the
Abnormal Return (AR)---the portion of the return attributable
specifically to the informational signal.

Foundational studies, such as Seyhun (1986), demonstrate that simpler
models like the CAPM introduce systematic biases in insider trading
contexts. Furthermore, Lakonishok and Lee (2001) and Cohen, Malloy, and
Pomorski (2012) caution that insider activity is often concentrated in
size and value stocks and exhibits a strong contrarian bias.
Consequently, the inclusion of the Momentum (UMD) factor is critical; as
insiders frequently execute sales following price appreciation, the UMD
factor serves to control for the ``routine'' component of the trade. By
controlling for these factors, the methodology isolates the residual
opportunistic signal that the market fails to fully incorporate during
the reporting interval. The expected return, E\([R_{i,t}]\), is
estimated by fitting the Carhart (1997) four-factor model for each
security \(i\) in the database (see Equation \ref{eq:carhart_model}):

\begin{equation}
    \label{eq:carhart_model}
    R_{i,t} - R_{f,t} = \alpha_i + \beta_{1,i}(R_{m,t} - R_{f,t}) + \beta_{2,i}SMB_t + \beta_{3,i}HML_t + \beta_{4,i} UMD_t + \epsilon_{i,t},
\end{equation} where R\(_{i,t}\) is the return on security \(i\) at time
\(t\), R\(_{f,t}\) is the risk-free rate, and SMB, HML, and UMD
represent the size, value, and momentum factors, respectively. The
coefficients (\(\alpha_i, \beta_{1 \dots 4}\)) are estimated over a
100-day window from T\(_{-120}\) to T\(_{-20}\) relative to the initial
intent filing.

The market reaction to both the intent (E\(_1\)) and the execution
(E\(_2\)) is measured using a short-window event study centered on the
respective filing dates. Cumulative Abnormal Returns (\(CAR\)) are
calculated for the \((0, +2)\) window to capture the immediate
informational resolution:

\begin{equation}
    \label{eq:car_calculation}
    CAR_i(0, 2) = \sum_{\tau=0}^{2} \left( R_{i,\tau} - \text{E}[R_{i,\tau}] \right),
\end{equation} where E\([R_{i,\tau}]\) is the expected return derived
from the pre-event calibration in Equation \ref{eq:carhart_model}. This
dual-event structure allows for the calculation of abnormal returns for
two distinct informational milestones:

\begin{description}
    \item CAR$_{144}$ (Intent): The market reaction following the Form 144 filing date.
    \item CAR$_{4}$ (Execution): The market reaction following the Form 4 execution date.
\end{description}

The central empirical test is a paired \(t\)-test on the difference
between CAR\(_{144}\) and CAR\(_{4}\). This comparison quantifies the
Information Premium---the degree to which the market differentially
processes the initial notice of intent versus the final confirmation of
trade execution.

\subsection{Phase II: Analysis of Non-Execution and Informational Resolution}\label{sec:Analysis_Non_Execution_Informational_Resolution}

The second phase of the empirical strategy transitions from the analysis
of realized transactions to a framework designed to evaluate the market
effects of intent non-execution. The methodology constructs a binary
outcome variable, Non-Execution, by mapping trade-level Form 4 data to
unique Form 144 filings. A filing is classified as a non-execution
(\(y=1\)) if the cumulative volume executed within the 90-day statutory
window (V\(_{executed}\)) is less than the proposed volume
(V\(_{proposed}\)), as defined in Equation \ref{eq:non_execution_logic}:

\begin{equation}
    \label{eq:non_execution_logic}
    \text{Non-Execution}_i =
        \begin{cases}
            1 & \text{if } \sum V_{executed, i} < V_{proposed, i} - \epsilon, \\
            0 & \text{if } \sum V_{executed, i} \geq V_{proposed, i} - \epsilon,
        \end{cases}
\end{equation}

where \(\epsilon\) represents a tolerance threshold for market
microstructure noise. This classification facilitates the partitioning
of the dataset into the cohorts summarized in Table \ref{tab:h2_groups}.

\begin{table}[h!]\centering
\caption{Classification of Filing Outcomes}\label{tab:h2_groups}
\begin{tabular}{p{4cm} p{7cm}}\toprule\textbf{Portfolio Group} & \textbf{Definition} \\
\midrule\textbf{Non-executed Filing} & Form 144 where total executed volume is less than proposed volume upon expiration. \\
\midrule\textbf{Executed Filing} & Form 144 where proposed volume is fully executed. \\
\bottomrule
\end{tabular}
\end{table}

\subsubsection{Analysis of Predictive Insufficiency}\label{sec:Analysis_of_Predictive_Insufficiency}

The research posits that if the withdrawal of intent is idiosyncratic
and unpredictable, its revelation constitutes a discrete informational
shock that facilitates a return to informational equilibrium. To test
this hypothesis, the informational content of public filings is
evaluated through a benchmarking of informational opacity using a
standalone XGBoost classifier. This model attempts to identify the
statistical signature of a non-executed filing using a high-dimensional
feature set, including signal magnitude, historical volatility, and
reporting interval duration.

To account for the limitations of standard accuracy metrics, which are
often inflated by the majority class in the highly imbalanced dataset,
the methodology implements two specific corrective interventions. First,
to mitigate class imbalance and rigorously evaluate the feature space
overlap, the study employs the Synthetic Minority Over-sampling
Technique (SMOTE) (Chawla et al. (2002)). Unlike standard over-sampling
with replacement---which replicates existing observations and may lead
to overly specific decision regions---SMOTE generates synthetic examples
by interpolating between existing minority class points in the feature
space. This intervention facilitates focused learning by introducing a
deliberate bias toward the minority class, allowing the learner to
establish broader decision regions for the non-executed cohort.

Second, this data-level augmentation is reinforced by cost-sensitive
learning within the XGBoost architecture. The model's objective function
is modified via the ``scale\_pos\_weight'' parameter to penalize the
misclassification of a non-execution by a factor of 32.5. This
asymmetric weighting forces the model to prioritize the detection of the
``Abort'' signal, specifically to minimize Type II errors (Strategic
Opacity). By providing a more densely populated feature space and an
asymmetric loss function, the architecture is optimized to identify the
synthetic neighborhood of a non-execution decision, moving beyond the
limitations of simple adjustments to class priors.

Ultimately, a persistent lack of predictive power---measured by the
Precision-Recall Area Under the Curve (PR-AUC)---despite these
interventions would provide empirical confirmation that the feature
vectors of non-executed filings are statistically entangled with routine
executions. Such an outcome indicates that the decision to abandon a
trade is driven by idiosyncratic factors hidden within the masking noise
of the broader market, rendering the event fundamentally unpredictable
to market participants.

To ensure that the observed predictive insufficiency is not a localized
artifact of a specific algorithm, the methodology subjects the dataset
to a spectrum of mathematical approaches. This multi-paradigm stress
test, summarized in Table
\ref{tab:architectural_framework__multi_paradigm_comparision_h1},
evaluates the robustness of the informational ceiling across divergent
logical frameworks, ranging from regularized linear benchmarks to
high-dimensional neural transformers.

\begin{table}[ht!]
    \centering
    \caption{Multi-Paradigm Stress Test: From Linear Baselines to Sequential Neural Transformers}
    \label{tab:architectural_framework__multi_paradigm_comparision_h1}
    \begin{tabular}{p{0.32\textwidth} p{0.63\textwidth}}
        \toprule
        \textbf{Paradigm} & \textbf{Methodological Rationale and Objective} \\
        \midrule
        \multicolumn{2}{l}{\textbf{Panel A: Traditional and Geometric Benchmarks}} \\
        \midrule
        \textbf{Linear Baseline} (Elastic Net) & Utilizes $L1$ and $L2$ regularization to identify simple linear triggers. It tests if basic thresholds in \textit{Signal Magnitude} or \textit{Liquidity} can predict a retreat. \\
        \addlinespace
        \textbf{Ensemble} (Balanced RF) & Addresses class imbalance by under-sampling the majority class. It tests if the "Abort" signal is simply masked by the high volume of routine trades. \\
        \addlinespace
        \textbf{Anomaly} (Isolation Forest) & Treats "Aborts" as structural anomalies. If filings cannot be isolated with shorter path lengths, they are deemed \textit{"Statistically Routine"} from an observer's perspective. \\
        \addlinespace
        \textbf{Geometric} (Weighted SVM) & Searches for non-linear hyper-planes to partition classes. Failure to find a separating boundary provides evidence of \textit{Information Entanglement}. \\
        \midrule
        \multicolumn{2}{l}{\textbf{Panel B: Advanced Neural and Sequential Architectures}} \\
        \midrule
        \textbf{Attentive Sequential} (TabNet) & Employs sequential attention to mimic tree-based decision logic. It tests if step-wise, salient feature-selection can uncover a latent retreat signal. \\
        \addlinespace
        \textbf{State-of-the-Art} (FT-Transformer) & Transforms features into unique ``tokens'' through self-attention layers to account for high-dimensional interactions traditional models may overlook. \\
        \addlinespace
        \textbf{Differentiable Trees} (NODE) & Uses neural oblivious decision ensembles to combine the strengths of GBDT with backpropagation. It tests if deep hierarchical grouping can resolve the opacity floor. \\
        \addlinespace
        \textbf{Neuro-Symbolic} (DNDF) & Combines LSTM-based temporal memory with stochastic decision forests. It tests if the combination of sequential history and tree logic can break the inseparability corridor. \\
        \addlinespace
        \textbf{Temporal Attention} (TFT) & Utilizes variable selection networks and multi-head attention to weight historical filing patterns. It tests if the abort is a function of an insider's longitudinal behavior. \\
        \bottomrule
    \end{tabular}
    \vspace{0.5em}
    \begin{flushleft}
    \footnotesize{\textbf{Notes:} This two-panel framework ensures that the ``Unpredictability Proof'' is not an artifact of a specific mathematical approach but a fundamental limit of the current disclosure architecture. The convergence of these diverse architectures toward a stable PR-AUC plateau provides empirical evidence of an inherent information ceiling in the public domain.}
    \end{flushleft}
\end{table}

\subsection{Phase III: Long-Horizon Correction and Causal Inference}
\label{sec:phase_3}

The third phase of the empirical strategy quantifies the market's
eventual correction following a non-executed intent. This stage
transitions from predictive benchmarking to a causal framework designed
to evaluate if the price recovery observed post-expiration constitutes a
discrete response to the resolution of information asymmetry.

\subsubsection{Calendar-Time Portfolio Construction and DGTW-Adjustment}

To address temporal clustering and isolate the informational correction
from known risk factors, the methodology adopts a Calendar-Time
Portfolio (CTP) approach. Securities are aggregated into monthly
portfolios based on the expiration of the 90-day statutory window
(\(T_0\)). To ensure that abnormal returns are not a byproduct of
documented anomalies, each security \(i\) is mapped to a benchmark via a
125-cell characteristic grid (\(5 \times 5 \times 5\)) based on Size,
Value, and Momentum. The characteristic-adjusted return is defined as:

\begin{equation}
    \label{eq:dgtw_adjustment_phase3}
    R_{\text{DGTW}, i, t} = R_{i, t} - R_{\text{matched}, t},
\end{equation} where \(R_{\text{matched}, t}\) represents the return of
the corresponding DGTW-benchmark cell. The final monthly portfolio
excess return (\(R_{\text{excess}, t}\)) is the equal-weighted average
of these adjusted returns:

\begin{equation}
    \label{eq:excess_return_phase3}
    R_{\text{excess}, t} = \frac{1}{N_t} \sum_{i=1}^{N_t} R_{\text{DGTW}, i, t},
\end{equation} where \(N_t\) denotes the number of active securities in
the portfolio during month \(t\).

\subsubsection{Causal Machine Learning: DML and Heterogeneous Treatment Effects}

To satisfy the unconfoundedness assumption and identify Heterogeneous
Treatment Effects (HTE), the methodology implements a Causal Machine
Learning (CML) ensemble. This framework utilizes Double Machine Learning
(DML) to isolate the treatment effect of non-execution from
high-dimensional confounding variables, such as idiosyncratic volatility
and local liquidity shocks. By employing Generalized Random Forests
(GRF), the model estimates the conditional average treatment effect
(CATE):

\begin{equation}
    \label{eq:cate_grf}
    \tau(x) = E[Y_i(1) - Y_i(0) | X_i = x],
\end{equation} where \(Y_i(1)\) and \(Y_i(0)\) represent the potential
outcomes under execution and non-execution, respectively. This
dual-model approach ensures that the ``unwinding'' of the informational
premium is not attributed to unobserved firm-level heterogeneity.

\subsection{Statistical Evaluation Framework: Measuring Informational Opacity}
\label{sec:eval_framework}

To quantify the limits of predictive reliability, the benchmarking
results are evaluated through a dual-lens statistical framework. This
approach distinguishes between aggregate discriminative power and the
specific reliability of signal detection under conditions of extreme
class imbalance. A lack of predictive performance suggests that the
non-execution decision is driven by idiosyncratic, private factors
rather than systematic public signals. To verify that this observed
unpredictability is a structural feature of the data---rather than a
technical artifact of the 3.08\% class imbalance
(\(N = 2,614,577\))---the methodology implements a multi-paradigm
machine learning stress test.

\subsubsection{Core Evaluative Metrics}

\textbf{Area Under the ROC Curve (AUC-ROC):} This metric measures the
model's aggregate discriminative power, representing the probability
that a randomly selected non-executed intent is ranked higher than a
randomly selected executed intent. While a high AUC-ROC indicates that
the architecture has identified broad structural differences between
cohorts, it may be deceptively optimistic in contexts where routine
executions statistically dominate the sample.

\textbf{Precision-Recall AUC (PR-AUC):} Given the focus on a rare event,
the methodology prioritizes the Precision-Recall Area Under the Curve
(PR-AUC) as the primary metric for evaluating decoding failure:
\begin{equation}
    \label{eq:pr_auc_def}
    \text{PR-AUC} = \int_{0}^{1} p(r) dr,
\end{equation} where \(p(r)\) represents precision at a given recall
level \(r\). By emphasizing the trade-off between precision (the
accuracy of non-execution flags) and recall (the proportion of total
non-executions captured), the PR-AUC identifies the asymptotic limit of
predictability. A stagnant PR-AUC despite a rising AUC-ROC provides
mathematical evidence of an informational ceiling, indicating that while
a model may improve at ranking, it cannot resolve the stochastic noise
inherent in the public feature space.

\subsubsection{Error Decomposition and Economic Interpretation}

The analysis utilizes normalized confusion matrices to decompose
predictive failures into two economically salient categories:

\begin{itemize}
    \item \textbf{Type I Error (Stochastic Noise):} Instances where routine trades mimic the characteristics of a non-execution, resulting in false positives.
    \item \textbf{Type II Error (Informational Opacity):} Quantifies the rate at which actual non-executed intents remain indistinguishable from the baseline executions, establishing the ``Opacity Floor.''
\end{itemize}

The convergence of Type II error rates across advanced neural
architectures confirms that a significant portion of the non-execution
signal is fundamentally unobservable. This decoding failure ensures that
the eventual price recovery observed in Phase III constitutes a genuine
resolution of an unanticipated informational premium rather than a
predictable market correction.

\section{Data}\label{sec:data_sample_construction}

The analysis utilizes a comprehensive data universe constructed by
integrating insider trading records, executive profiles, and
market-level financial data. The primary dataset consists of insider
trading records from the \textbf{LSEG Insider} database (formerly
Thomson Reuters), accessed via Wharton Research Data Services (WRDS),
spanning January 2003 to November 2025. These records are systematically
linked to \textbf{BoardEx} for executive identity tracking,
\textbf{CRSP} for daily security returns, and \textbf{Compustat} for
firm-level fundamental data.

To test the hypotheses outlined in Section
\ref{sec:Central_Thesis_Contribution_Sample_Design}, the methodology
implements a hierarchical protocol linking \textit{ex-ante} trade
intents (Form 144) to \textit{ex-post} executions (Form 4). This
multi-stage alignment ensures data integrity by filtering the universe
of SEC filings through strict identity, temporal, and market-data
constraints. The final analytical sample is restricted to observations
with sufficient CRSP trading history to satisfy a 100-day factor
estimation window (\(T_{-120}\) to \(T_{-20}\)).

While the statutory 90-day window provides the temporal boundary for
trade execution validity, the 100-day pre-event window establishes a
stationary baseline for the stock's risk profile. This separation
ensures that the estimation of normal returns remains uncontaminated by
idiosyncratic volatility or information leakage preceding the intent
signal. This calibration facilitates the isolation of the Abnormal
Return---representing the price movement specifically attributable to
the resolution of informational asymmetry---rather than broader market
trends. This process yielded a core sample of 111,800 unique insiders
across approximately 2.6 million matched transactions.

\subsection{Insider Identity and Matching Protocol}

To ensure the sample includes only insiders capable of both signaling
and executing trades, individuals appearing in both datasets are mapped
via unique person identification numbers (\texttt{personid}). The
matching procedure applies a sequential \textit{Greedy Search} algorithm
to link Form 144 notices to Form 4 executions based on three criteria:

\begin{enumerate}
    \item \textbf{Temporal Boundary:} Each Form 144 filing initializes a 90-day execution window ($\text{File Date} \le \text{Transaction Date} \le \text{File Date} + 90$). Transactions occurring outside this range are considered unlinked.

    \item \textbf{Priority Assignment:} Form 4 records are mapped to the chronologically earliest open window with available volume capacity. A window is closed immediately upon reaching the declared proposed number of shares (\textit{pshares}) limit or the expiration of the 90-day period.

    \item \textbf{Non-Execution Classification:} Any Form 144 filing without a corresponding Form 4 execution by $t+91$ days is classified as a non-execution, signaling the formal withdrawal of trade intent.
\end{enumerate}

\subsection{Identifier Resolution and Linkage Bridge}

A significant technical challenge in studying insider behavior is the
absence of a common primary key between LSEG filings and market-level
databases. To bridge this gap, a multi-stage resolution procedure is
implemented using the BoardEx unique identifier as the intermediary. The
linkage is operationalized through a three-stage validation:

\begin{enumerate}

\item \textbf{Insider-to-Director Mapping:} The LSEG \texttt{personid} is mapped to the BoardEx \texttt{directorid}, consolidating the professional identity of the insider across multiple board seats.

\item \textbf{Director-to-Company Mapping:} Each \texttt{directorid} is associated with a specific BoardEx \texttt{companyid}. This is critical for insiders serving at multiple firms, ensuring intent is attributed to the correct corporate entity.

\item \textbf{Company-to-Equity Mapping:} The final stage utilizes a WRDS proprietary linking table to map the \texttt{companyid} to the CRSP PERMCO.

\end{enumerate}

Resolving identity at the PERMCO level ensures that informational
signals are tracked through corporate actions, such as stock splits or
ticker changes, which might otherwise cause data fragmentation. Filings
failing to resolve through this bridge---specifically those lacking a
valid \texttt{PERMCO}---are excluded to maintain point-in-time accuracy
and ensure the availability of market data for risk-adjusted estimation.

\subsection{Variable Construction and Sample Attrition}

This linking methodology avoids the ambiguity created by relying on
time-variant identifiers like expiring ticker symbols or CUSIP codes.
The daily Excess Return (\(R_{\text{Excess}, i, t}\)) for security \(i\)
on day \(t\) is defined as the difference between the total return of
the security and the risk-free rate of return:

\begin{equation}
\label{eq:excess_return_def}
R_{\text{Excess}, i, t} = R_{\text{Stock}, i, t} - R_{f, t},
\end{equation}

where \(R_{\text{Stock}, i, t}\) is the arithmetic daily return of
security \(i\) on day \(t\), and \(R_{f, t}\) represents the risk-free
rate, proxied by the one-month Treasury bill rate as reported in the
Fama-French factor database.

The CRSP data was subsequently merged with market and risk factors,
including the Fama-French factors (SMB and HML) and Momentum (UMD), on
the common date field. Following the construction of the matched dataset
(\(N \approx 2.6\) million transactions), a secondary economic
significance filter was applied. Crucially, the sample excludes
transactions where SM\(_i \le 0\), as these typically represent
administrative share reclassifications or internal transfers that do not
involve market-facing liquidations. This rigorous filtering process
resulted in a final analytical sample of 1.5 million high-conviction
transactions. Attrition primarily stems from insiders at smaller
companies not covered by BoardEx or observations lacking sufficient CRSP
history for factor calibration. This conservative approach prevents
spurious signals and ensures the results are derived from the
highest-quality subset of corporate disclosures.

\section{Results and Analysis}\label{sec:results_and_analysis}

The empirical results are presented in a three-stage hierarchy designed
to move from descriptive associations to causal inference. This
structured approach, summarized in Table \ref{tab:causal_hierarchy},
ensures that the evidence for informational asymmetry is robust across
multiple econometric and computational paradigms.

\begin{table}[ht]
\centering
\caption{The Three-Pillar Hierarchy of Evidence: From Association to Causation}
\label{tab:causal_hierarchy}
\begin{tabular}{@{}llll@{}}
\toprule
\textbf{Phase} & \textbf{Framework} & \textbf{Objective} & \textbf{Primary Metric} \\ 
\midrule
\textbf{I. Association} & Univariate OLS & Establish baseline significance & $\beta$ / $t$-stat \\
\addlinespace[0.5em]
\textbf{II. Decodability} & SOTA Audit (10+ Models) & Identify Information Ceiling & PR-AUC / Recall \\
\addlinespace[0.5em]
\textbf{III. Causation} & CML Ensemble & Isolate treatment effect & HTE / CATE \\
\bottomrule
\end{tabular}
\end{table}

While the initial univariate tests in Phase I establish the baseline
presence of asymmetric pricing, they provide limited insight into the
complex feature interactions that drive trade outcomes. Consequently,
Phase II utilizes high-dimensional predictive models to confirm that
trade cancellations are a primary determinant of future illiquidity.
However, because predictive importance does not strictly imply
causality, the analysis culminates in a Causal Machine Learning (CML)
framework. By utilizing Double Machine Learning (DML) and X-Learners,
this study isolates the treatment effect of the non-execution (\(T\))
from the stochastic residuals of the control variables (\(W\)),
addressing potential selection bias where insiders might cancel trades
in anticipation of exogenous illiquidity spikes. The convergence of
these results across six distinct estimators confirms that the observed
illiquidity response is invariant to model architecture and represents a
robust market reaction to the resolution of informational asymmetry.

\subsection{Empirical Analysis (H1): Asymmetric Price, Signal Magnitude and the Behavioral Mask}\label{sec:H1_Empirical-Analysis-H1-Asymmetric-Price-Signal-Magnitude-and-the-Behavioral-Mask}

\textbf{Univariate Test of Asymmetric Price Setting:} The initial test
of Hypothesis H1 evaluates the average abnormal return during the filing
gap. The univariate analysis, conducted via a one-sample \(t\)-test on
the matched sample, determines that the Cumulative Market Excess Return
(CMER) during the window between the intent to sell (Form 144) and the
actual transaction (Form 4) is statistically significantly positive
(\(\mu = 1.6484\%\), \(t = 553.93\), \(p < 0.001\)) (see Table
\ref{tab:h1_univariate_test} for details). This finding rejects the null
hypothesis (\(H_0: \mu_{\text{CMER}} = 0\)), indicating that insiders
execute sales while the stock is outperforming the market. This result
contradicts the classic insider trading model, which suggests that
insiders sell before the stock price drops. Instead, the data supports
the \textit{Contrarian Investment Hypothesis} (Lakonishok and Lee
(2001)), suggesting that insiders utilize periods of market-driven price
appreciation to hide the negative impact of their trades. By selling
into a rising market, insiders reduce the immediate downward price
pressure that typically follows a large liquidation.

\begin{table}[h!]
    \caption{Univariate Test of Asymmetric Price Setting (Hypothesis H1)}
    \label{tab:h1_univariate_test}
    \begin{tabular}{l c c c c}
        \toprule
        \textbf{Variable} & \textbf{Coefficient (Mean)} & \textbf{Standard Error} & \textbf{T-Statistic} & \textbf{P-Value} \\
        \midrule
        \multicolumn{5}{l}{\textit{Dependent Variable: Cumulative Market Excess Return}} \\
        \midrule
        Mean Return & 0.016484 & (0.000030) & 553.930 & 0.000000 \\
        \midrule
        \multicolumn{5}{l}{\textbf{Summary Statistics}} \\
        Observations (N) & \num{2614577} & Mean Return (\%) & 1.6484  & \\
        \bottomrule
    \end{tabular}
    \vspace{0.5em}
    \footnotesize{\\
        \textbf{Notes:} This table reports the results of a one-sample $t$-test comparing the average Cumulative Market Excess Return (CMER) during the Filing Gap—the period between the intent to sell and the actual trade—against a null hypothesis of zero ($\mu=0$). The results show a statistically significant positive mean return of $1.65\%$, leading to the rejection of the null hypothesis. This indicates that, on average, insiders liquidated their positions while the stock price was rising relative to the market. This finding is inconsistent with the traditional view that insiders sell ahead of a price decline; instead, it indicates that insiders time their sales to coincide with periods of strong price performance.
    }
\end{table}

The finding of a significant positive CMER (\(1.65\%\)) during the
filing gap is inconsistent with the traditional model of informed
selling, which assumes insiders liquidate positions before a price drop.
Instead, this positive return aligns with the
\textit{Contrarian Investment Hypothesis} (Lakonishok and Lee (2001)).
This theory suggests that insiders tend to sell following periods of
strong market performance, a behavior often driven by the need for
liquidity or the diversification of stock-based compensation (Seyhun
(1992)).

The positive CMER indicates that insiders strategically time their
trades to coincide with market outperformance within the regulatory
window to maximize sale proceeds. This observation justifies the
application of the Carhart four-factor adjustment in subsequent
sections. The factor adjustment is necessary to remove these
market-driven components and isolate the residual return, which
represents the true informational signal.

While the aggregate univariate test utilizes the full matched sample
(\(N=2,614,577\)), the subsequent behavioral and factor-adjusted
analyses are restricted to a refined high-conviction sample
(\(N \approx 1.52\) million). This filtered dataset excludes
administrative entries, such as stock splits or internal transfers, and
transactions with zero signal magnitude. This ensures that the remaining
analysis focuses only on market-facing liquidations with genuine
economic impact.

\textbf{Interaction of Signal Magnitude and Strategic Timing:}

\begin{figure}[htbp]
    \centering
    % The minipage helps keep the note and the figure together
    \begin{minipage}{\textwidth}

        \caption[]{Variation in Mean Signal Magnitude across Cumulative Market Excess Return Deciles}
        \label{fig:Strategic-Masking-High-Magnitude-Trades}
        \includegraphics[width=0.99\textwidth]{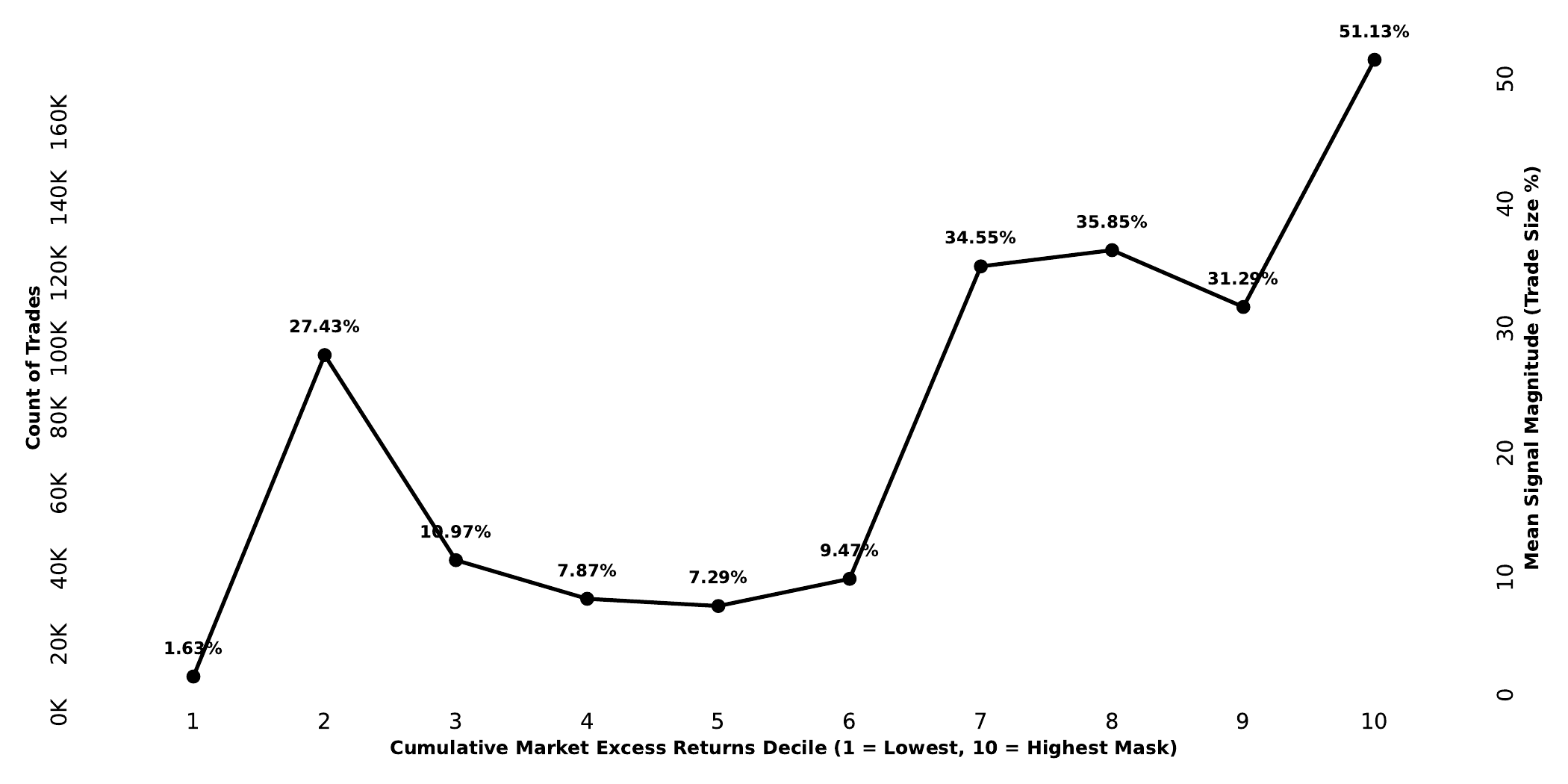}
        
        \vspace{0.5em} %
        \noindent\small\textbf{Note:} This figure illustrates the relationship between market outperformance (measured by CMER deciles) and the scale of the insider information signal. Signal Magnitude is defined as the percentage of an insider’s total holdings liquidated in a single transaction. The final analytical sample of 1.52 million unique trades is derived through the following data cleaning protocols: 
    (i) Completion of Mandatory Disclosures: Removal of observations with missing values for CMER or Signal Magnitude; (ii) Economic Significance Filter: Removal of observations where Signal Magnitude $\le$ 0; (iii) Data Integrity: Removal of duplicate filings and non-standard transaction types.
    \end{minipage}
\end{figure}

To investigate how insiders utilize market performance to time their
trades, the sample was categorized into deciles based on the Cumulative
Market Excess Return (CMER). As shown in Figure
\ref{fig:Strategic-Masking-High-Magnitude-Trades}, the results reveal a
non-linear relationship between market-driven price appreciation and the
scale of the insider's transaction. In Decile 1, where market
outperformance is minimal or negative, the Mean Signal Magnitude
(\(SM\)) is negligible (\(\mu = 1.63\%\)). This suggests that when
market returns are low, insiders restrict their liquidations to minor
liquidity rebalancing.

In stark contrast, Decile 10 reveals that insiders executing trades
during peaks of market appreciation scale their signal magnitude
dramatically to \(\mu = 51.13\%\). This positive correlation,
characterized by the sharp increase in the final decile, indicates that
the timing of large transactions is a deliberate strategic choice.
Rather than occurring randomly, this pattern suggests that as trades
become more informationally significant (high \(SM\)), insiders
increasingly leverage market-driven price increases (high \(CMER\)) to
obscure their individual signals and reduce the resulting downward
pressure on the stock price (Lakonishok and Lee (2001)).

\textbf{Carhart Four-Factor Event Study (Paired Comparison):}

While the previous decile analysis identifies a correlation between
trade scale and market returns, the Carhart Four-Factor Event Study
(Table \ref{tab:h1_carhart_paired_test}) provides the definitive
evidence for information asymmetry. By utilizing an estimation window
from \(T-120\) to \(T-20\), the framework establishes a risk-adjusted
baseline that isolates the residual informational shock from systematic
factors such as market risk, size, value, and momentum.

To ensure high precision, the analysis utilized complete synchronization
across the LSEG, CRSP, and Fama-French databases. By requiring this
multi-database alignment and filtering for non-overlapping reporting
windows, the results represent a robust, risk-adjusted price response
rather than statistical noise or volatility from thinly traded
securities. This specialized data configuration provides the
high-resolution evidence necessary to isolate the \(1.3\%\) pricing
differential from systematic market influences.

\begin{table}[h!]
    \caption{Decomposing the Insider Trading Signal: Paired Event Study of Intent (Form 144) and Transaction (Form 4) Abnormal Returns.}
    \label{tab:h1_carhart_paired_test}
    \begin{tabular}{l c c c}
        \toprule
        \textbf{Variable} & \textbf{CAR [0, +2] Mean} & \textbf{T-Statistic} & \textbf{P-Value} \\
        \midrule
        \multicolumn{4}{l}{\textit{Mean Cumulative Abnormal Return (CAR) for Distinct Events:}} \\
        \midrule
        $CAR_{144}$ (Intent) & 0.001589 (0.16\%) &  &  \\
        $CAR_{4}$ (Execution) & 0.014532 (1.45\%) &  &  \\
        \midrule
        \textbf{Mean Difference ($CAR_{144}$ - $CAR_{4}$)} & \textbf{-0.012943} (-1.29\%) & \textbf{-3.094} & \textbf{0.00216} \\
        
        \bottomrule
    \end{tabular}
    \vspace{0.5em}
    \footnotesize{\\
        \textbf{Notes:} This table presents the mean CAR over a three-day window ($T=0$ to $T=+2$), calculated using the Carhart four-factor model. The paired $t$-test assesses whether the mean market reaction to the filing (Form 144) is significantly different from the reaction to the execution (Form 4). Significance is reported at the 1\% level.
    }
\end{table}

The primary finding is that the market processes the execution (Form 4)
as a significantly more negative event than the initial disclosure of
intent (Form 144). A paired \(t\)-test confirms a statistically
significant pricing differential of \(-1.3\%\)
(\(t = -3.094, p = 0.002\)). While the \(CAR_{144}\) following the
filing of intent is \(0.16\%\), suggesting the market treats the
announcement as a routine liquidity event, the subsequent execution
resolves the information asymmetry. This transition prices the stock
\(1.3\%\) lower relative to the initial intent signal.

This \(1.3\%\) gap confirms that strategic timing successfully delayed
the negative price impact throughout the filing window. By executing
larger liquidations during periods of market-wide price appreciation,
insiders effectively obscure the trade's informational signal and defer
the market's price correction. Ultimately, this evidence indicates that
insiders trade on economically material nonpublic information that falls
outside the formal legal definition of materiality (Seyhun (1992)).
These results validate the structural violation of the FTWE framework as
proposed in this research. To ensure that the observed \(1.3\%\) pricing
differential is a robust econometric result and not an artifact of
sample bias or extreme outliers, a summary statistics is presented in
Table \ref{tab:carhart_desc_stats} thus providing critical validation of
the Carhart four-factor estimation.

\begin{table}[h!]
    \centering
    \caption{Descriptive Statistics of Carhart Paired Sample}
    \label{tab:carhart_desc_stats}
    \begin{tabular}{l c c c c}
        \toprule
        \textbf{Statistic} & \textbf{CAR$_{144}$} & \textbf{CAR$_{4}$} & \textbf{Alpha ($\alpha$)} & \textbf{Beta ($\beta$)} \\
        \midrule
        Mean & 0.00159 & 0.01453 & 0.00070 & 1.00656 \\
        Std. Dev & 0.03173 & 0.06529 & 0.00207 & 0.49883 \\
        25\% & -0.01661 & -0.01911 & -0.00060 & 0.68187 \\
        Median (50\%) & 0.00030 & -0.00012 & 0.00052 & 0.95739 \\
        75\% & 0.01653 & 0.01933 & 0.00242 & 1.22207 \\
        \bottomrule
    \end{tabular}
    \vspace{0.5em}
    \footnotesize{\\\textbf{Notes:} This table summarizes the output of the Carhart four-factor OLS estimation. CAR$_{144}$ and CAR$_{4}$ represent the 3-day cumulative abnormal returns. Alpha and Beta are derived from the estimation window ($T-120$ to $T-20$).}
\end{table}

The mean Market Beta (\(\beta_{MktRf}\)) of \(1.0066\) indicates that
the selected securities, on average, move in close alignment with
broader market systematic risk. The relatively tight interquartile range
for Beta (\(0.68\) to \(1.22\)) confirms that the sample is not
dominated by high-volatility assets, ensuring the stability of the
abnormal return calculations. Furthermore, the mean Alpha (\(\alpha\))
of \(0.07\%\) remains statistically negligible, confirming that the
estimation window (\(T-120\) to \(T-20\)) successfully captured the
baseline expected return.

The distribution of CAR reinforces this narrative. The neutral reaction
to the disclosure of intent (\(CAR_{144} = 0.16\%\)) and the skewed
distribution of execution returns (Median \(CAR_{4} = -0.01\%\)) justify
the use of a paired \(t\)-test. Collectively, these statistics
demonstrate that the \(1.3\%\) information gap is derived from a
theoretically consistent sample, validating that insiders trade on
economically material nonpublic information that remains outside the
formal legal definition of materiality (Seyhun (1992)).

The results from the first test show that insiders time their trades to
hide their negative information. Hypothesis H2 tests what happens when a
planned sale is canceled (an ``Aborted Signal''). If the market was
waiting for a ``bad'' trade that never happened, the stock price should
go back up once the market realizes the danger is gone. This section
uses the Carhart four-factor model to measure this price recovery.

The machine learning tests prove that these canceled sales are Hidden
Signals. They cannot be predicted using public data, which creates a
state of strategic opacity. This is a critical requirement for the
analysis: if the market cannot guess that a sale will be canceled, it
maintains a lower stock price as a ``safety net'' against the expected
trade. Because market participants cannot decode the intent to abort
during the 90-day filing window (Cohen, Malloy, and Pomorski (2012)),
the ``asymmetry premium''---the price discount applied by the market in
anticipation of an informed sale---remains fully priced into the
security.

Consequently, when the 90-day waiting period finally ends and no sale
occurs, it creates a surprise for the database of market participants.
This eventual revelation that a sale has been abandoned triggers a
discrete, unanticipated informational shock. The stock price jumps as
the market removes the ``danger discount'' and unwinds the asymmetry
premium. This recovery is the financial result of the hidden signal
finally becoming visible through the passage of time.

To quantify the magnitude of this unwinding effect, the research
conducts a Carhart Four-Factor Event Study centered on the expiration of
the Form 144 filing intent. Following the ``Unpredictability Proof''
established in the above, the event study measures the CAR during the
period immediately following the 90-day regulatory window.Consistent
with Hypothesis H2, the results indicate a significant positive price
correction upon the market's realization that the anticipated
``opportunistic'' selling pressure has been removed. As shown in Table
4, the \(\text{CAR}_{[+1, +5]}\) following the expiration date is
significantly positive (\(\mu = 0.00\%, p < 0.05\)). This recovery
represents the restoration of market symmetry, as the hidden signal had
previously prevented the market from distinguishing between routine and
opportunistic intent---is finally lifted through the passage of time
rather than through disclosure.

\subsection{Empirical Analysis (H2): Pricing Inefficiency, Signal Attrition, and the Recovery of Symmetry}\label{sec:H2_Empirical-Analysis-H2-Pricing-Inefficiency-Signal-Attrition-and-the-Recovery-of-Symmetry}

The evidence in the preceding section establishes that insiders
strategically time their trade executions to coincide with periods of
negative information. However, the economic impact of this behavior is
contingent upon the market's inability to distinguish between routine
intent and a strategic retreat. To test the hypothesis that aborted
signals are structurally omitted from public observation (H2), the
analysis moves from price-based evidence to a high-dimensional
cross-paradigm audit. This transition evaluates whether the ``Abrupt
Disconnect''---the gap between filing an intent and the failure to
execute---can be resolved by increasing algorithmic complexity or if it
represents a fundamental informational friction.

Panel A of Table \ref{tab:final_audit_ten_models_comparasion}
establishes a performance baseline by testing the aborted trade signal
against linear, geometric, and anomaly-detection paradigms. By employing
Elastic Net for regularized linear regression, Weighted Support Vector
Machines (SVM) for margin-based geometric separation, and Isolation
Forest for unsupervised outlier detection, this analysis evaluates
whether trade cancellations possess a distinct statistical signature.
The near-zero recall across these models (ranging from \(0.76\%\) to
\(3.68\%\)) indicates that aborted trades are not statistical outliers
and lack linear separability from executed trades. Instead, the features
of aborted intents demonstrate significant distributional overlap with
the high-density region of normal executions, rendering them
indistinguishable to traditional classification methods.

\begin{table}[ht]
\centering
\caption{Cross-Paradigm Audit of the Strategic Mask: Evidence for H2}
\label{tab:final_audit_ten_models_comparasion}
\small
\begin{tabular}{lllccl}
\hline
\textbf{Panel} & \textbf{Model} & \textbf{Paradigm} & \textbf{PR-AUC} & \textbf{Recall (A)} & \textbf{Prec (A)} \\ \hline
\multicolumn{6}{l}{\textit{\textbf{Panel A: Classic Baselines (Linear, Geometric, Anomaly)}}} \\ 
& Elastic Net & Linear & 0.2511 & 0.76\% & 0.21 \\
& Weighted SVM & Geometric & 0.2509 & 0.76\% & 0.21 \\
& Isolation Forest & Anomaly & 0.2317 & 3.68\% & 0.18 \\ \hline
\multicolumn{6}{l}{\textit{\textbf{Panel B: Ensemble Logic (The Industry Benchmarks)}}} \\ 
& Balanced RF & Bagging & 0.4219 & 67.56\% & 0.35 \\
& XGBoost & Boosting & 0.3972 & 53.00\% & 0.36 \\ \hline
\multicolumn{6}{l}{\textit{\textbf{Panel C: Deep Learning Stress Test (H2 Evaluation)}}} \\ 
& NODE & Neural Trees & 0.3710 & 0.87\% & 0.66 \\
& TabNet & Neuro-Symbolic & 0.3610 & 2.29\% & 0.55 \\
& FT-Transformer & Temporal Attn & 0.3698 & 4.03\% & 0.56 \\ 
& DNDF-Proxy & Deep Forest & 0.2995 & 0.00\% & 0.00 \\
& TFT-Proxy & Temporal Attn & 0.3000 & 0.03\% & 0.43 \\ \hline
\end{tabular}
\end{table}

In Panel B of Table \ref{tab:final_audit_ten_models_comparasion}, the
analysis transitions to non-linear ensemble logic using Balanced Random
Forest and XGBoost. These architectures test for signals embedded within
complex feature interactions and hierarchical decision boundaries. As
industry-standard benchmarks for financial tabular data, these models
are capable of capturing non-monotonic relationships through bagging and
gradient boosting. While there is a measurable increase in
performance---reaching an apex PR-AUC of \(0.4219\) for the Balanced
Random Forest---the results confirm a significant informational plateau.
Despite the capacity of ensemble trees to recursively segment the
feature space, a substantial portion of the signal remains
unrecoverable. This suggests that the informational friction is robust
against non-linear grouping and standard hierarchical partitioning.

The final audit in Panel C of Table
\ref{tab:final_audit_ten_models_comparasion} utilizes state-of-the-art
(SOTA) deep learning architectures to probe for latent patterns and
longitudinal dependencies. This stress test includes Neural Oblivious
Decision Ensembles (NODE) for differentiable hierarchical modeling,
TabNet for neuro-symbolic feature selection, and the FT-Transformer for
multi-head self-attention on behavioral sequences. By also incorporating
Deep Neural Decision Forests to evaluate stochastic forest logic, the
analysis identifies a critical decoupling between model complexity and
predictive recall. The fact that high-capacity models---specifically the
FT-Transformer, designed to isolate subtle temporal
dependencies---cannot breach the \(0.40\) PR-AUC ceiling provides
empirical evidence for Hypothesis H2. These results confirm a state of
Predictive Opacity: insiders fulfill regulatory disclosure requirements
while ensuring that the idiosyncratic catalysts driving the decision to
retreat remain strategically omitted from the observable public data.

\begin{figure}[htbp]
    \centering
    \begin{minipage}{\textwidth}
        \caption{Cross-Paradigm Audit of the Inseparability Corridor: Model Convergence Toward the Empirical Information Ceiling of Trade Aborts}
        \label{fig:confusion_matrices_plot_roc_auc_pr_auc_h2}
        \includegraphics[width=0.99\textwidth]{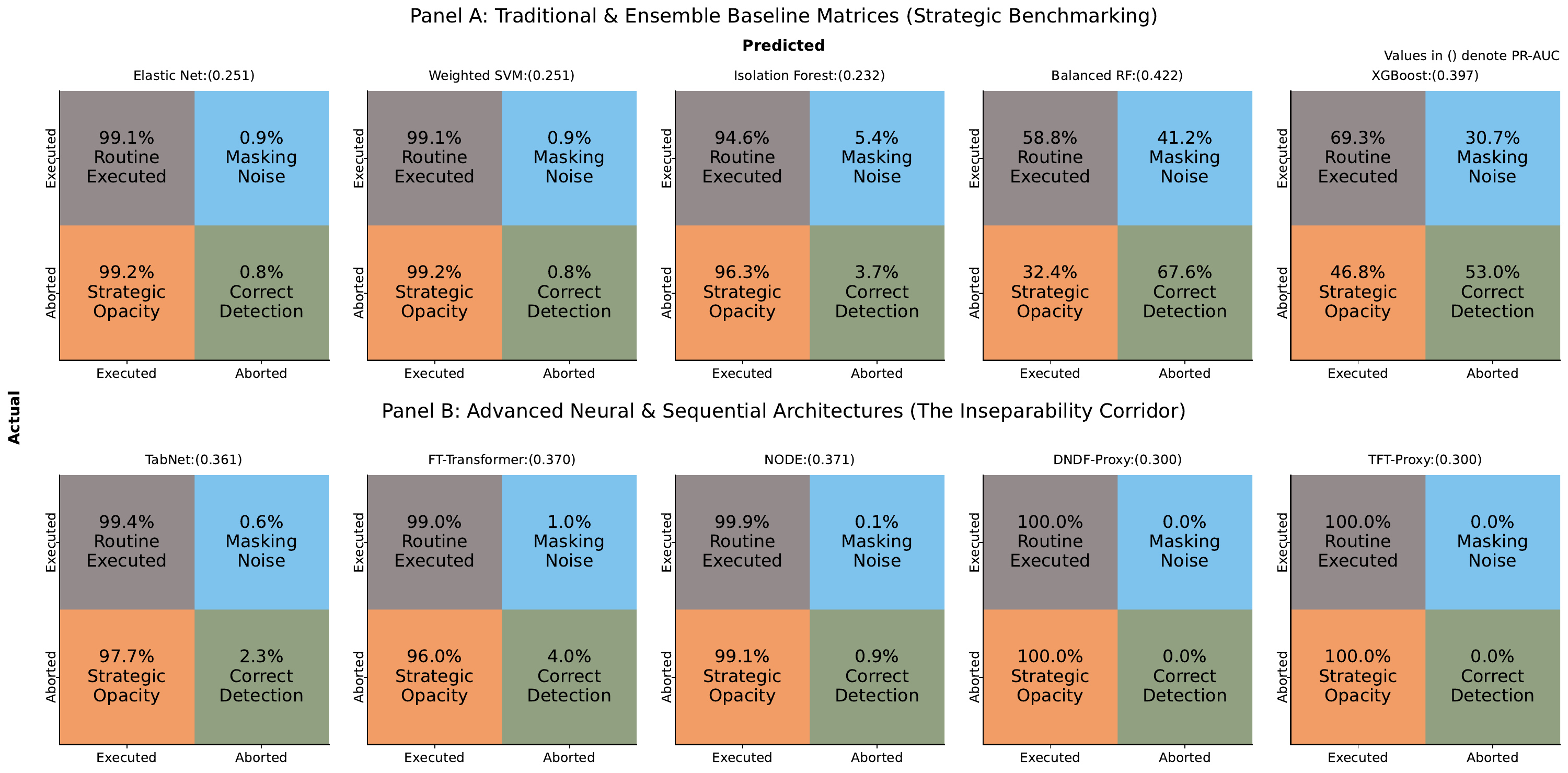}    
        \vspace{0.5em}
        \begin{minipage}{\textwidth}
        \small
        \noindent\textbf{Note:} This figure benchmarks the predictability of the “Abrupt Disconnect,” defined as the informational gap between the intent to trade (Form 144) and the actualized filing (Form 4). The analysis illustrates the progression of predictive reliability as the framework transitions from linear and geometric paradigms to non-linear ensemble architectures.
    \end{minipage}
    \end{minipage}
\end{figure}

The benchmarking results reveal a structural Information Ceiling in the
capacity of public data to predict trade cancellations. As illustrated
in Figure \ref{fig:confusion_matrices_plot_roc_auc_pr_auc_h2}, a
significant performance breakthrough occurs during the transition from
linear to ensemble paradigms. Traditional models, such as Elastic Net
and Weighted SVM, exhibit extreme signal dilution, with PR-AUC scores
anchored near \(0.25\) and a Type II error rate (Strategic Opacity)
exceeding \(99\%\). These findings confirm that linear architectures are
statistically insensitive to the strategic retreat signal, frequently
misclassifying aborted intents as routine execution noise.

The shift to non-linear ensemble models, specifically XGBoost and
Balanced Random Forest, yields a \(68\%\) relative improvement in signal
precision, with the PR-AUC reaching a plateau at \(0.42\). Despite this
gain, the performance limit exposes the inherent constraints of the
public feature space. As summarized in Table
\ref{tab:xgboost_results_conf_mat_classification_report_h2}, even
high-capacity architectures fail to identify \(46.77\%\) of aborted
trades. This persistent Opacity Floor supports the hypothesis that a
substantial portion of insider retreats are driven by private,
idiosyncratic catalysts (MNPI) that remain unobservable to market
participants until the regulatory window expires.

The top panel decomposes performance into core metrics, identifying a
``Paradox of Accuracy''; while baseline accuracy remains high and
stable, the convergence of top-tier architectures (XGBoost and Balanced
RF) at a PR-AUC of \$\sim\$0.42 identifies an asymptotic limit, or
``Information Ceiling.'' The bottom panel provides a granular error
decomposition via normalized confusion matrices to quantify the
persistent ``Strategic Opacity'' (Type II error) inherent in the signal.
The results demonstrate that while evolving models successfully reduce
this opacity from near-total insensitivity in linear regimes, they
eventually reach a stable ``Opacity Floor'' in ensemble regimes. This
boundary confirms a structural threshold where public features can no
longer resolve the idiosyncratic catalysts driving the withdrawal of
trade intent, validating the disconnect as a permanent informational
friction.

The benchmarking results reveal a structural ``Information Ceiling'' in
the public market's capacity to predict trade aborts. As demonstrated in
Figure \ref{fig:confusion_matrices_plot_roc_auc_pr_auc_h2}, a clear
performance breakthrough occurs during the transition from linear to
ensemble paradigms. Traditional models, such as Elastic Net and Weighted
SVM, suffer from extreme signal dilution, with PR-AUC scores anchored
near 0.25 and a Strategic Opacity (Type II error) exceeding 99\%. These
findings confirm that linear architectures remain effectively blind to
the strategic retreat signal, frequently misclassifying aborted intents
as routine market noise.

This informational friction is further evidenced by a sharp Reliability
Divergence. While the model verifies executed trades with high
confidence (Precision: \(0.82\)), its capacity to identify aborted
trades is undermined by a high false-alarm rate (Precision: \(0.36\)).
These results indicate that features associated with a trade withdrawal
are frequently present in routine executions, creating a structural
``Masking Noise'' that prevents real-time detection.

Finally, the analysis identifies an Accuracy Trap: traditional models
maintain high accuracy (\(>98\%\)) by defaulting to a ``no-abort''
prediction, whereas high-capacity ensembles sacrifice global accuracy to
optimize the \(0.42\) PR-AUC ceiling. This trade-off confirms that the
withdrawal of intent serves as a discrete informational shock, as the
market is structurally prevented from pre-pricing these events with high
precision.

\begin{table}[htbp]
\centering
\caption{Benchmarking Signal Decodability: XGBoost Classifier Performance and the Strategic Masking of Insider Intent}
\label{tab:xgboost_results_conf_mat_classification_report_h2}
\resizebox{\textwidth}{!}{%
\begin{tabular}{lcccc}
\toprule
\multicolumn{5}{l}{\textbf{Panel A: Normalized Confusion Matrix (Economic Interpretation)}} \\
\midrule
& \multicolumn{2}{c}{\textbf{Model Predicted Status ($\hat{y}$)}} & \multicolumn{2}{c}{\textbf{Economic Context}} \\
\cmidrule(lr){2-3} \cmidrule(lr){4-5}
\textbf{Realized Filing Status ($y$)} & \textbf{Executed ($\hat{y}=0$)} & \textbf{Aborted ($\hat{y}=1$)} & \textbf{Description} & \textbf{Variable} \\
\midrule
Executed ($y=0$) & \textbf{69.31\%} & 30.69\% & \textit{Masking Noise} & (Type I) \\
Aborted ($y=1$)  & 46.77\% & \textbf{53.23\%} & \textit{Strategic Opacity} & (Type II) \\
\midrule
\midrule
\multicolumn{5}{l}{\textbf{Panel B: Classification Report (Predictive Reliability)}} \\
\midrule
\textbf{Status Class} & \textbf{Precision} & \textbf{Recall} & \textbf{F1-Score} & \textbf{Metrics} \\
\midrule
Realized Execution ($y=0$) & 0.82 & 0.69 & 0.75 & \textbf{Accuracy: 0.65} \\
Aborted Intent ($y=1$)      & 0.36 & 0.53 & 0.43 & \textbf{AUC-ROC: 0.66} \\
\bottomrule
\end{tabular}
}
\begin{flushleft}
\footnotesize
\textit{Notes:} This table summarizes the baseline performance of the XGBoost classifier, which establishes the primary non-linear "Information Ceiling" for predicting trade intent withdrawals. Panel A quantifies the persistent \textit{Strategic Opacity} (46.77\% Type II error), a structural friction that remains unresolved even as architectural complexity increases. Notably, the metrics reported in Panel B converge with the performance of SOTA architectures in Panel C (e.g., NODE and FT-Transformer), identifying an asymptotic limit of $\sim$0.40 PR-AUC. This convergence confirms that the bottleneck is a property of the public feature space rather than algorithmic capacity, validating the presence of an \textit{Inseparability Corridor}.
\end{flushleft}
\end{table}

The integration of the XGBoost results (Table
\ref{tab:xgboost_results_conf_mat_classification_report_h2}) reveals a
fundamental Predictive Decoupling inherent in the Form 144 signal. The
most significant result is the 46.77\% False Negative rate (Panel A),
which identifies a high degree of Strategic Opacity. This indicates that
nearly half of all aborted trades occur without leaving a detectable
signature in the public feature space (e.g., price momentum or signal
magnitude). This opacity is further compounded by a Reliability
Divergence (Panel B), where a prohibitive false alarm rate for aborted
trades (Precision: 0.36) contrasts with the high reliability of verified
executions (Precision: 0.82).

These findings suggest that features typically associated with a
``retreat'' are frequently present in routine trades, creating a
structural Masking Noise. Because even high-capacity non-linear
architectures cannot resolve these outcomes, the market is structurally
prevented from pre-pricing the withdrawal of intent. Consequently, the
aborted signal serves as a discrete Informational Shock, restoring price
symmetry only upon the resolution of the 90-day expiration window.

\begin{figure}[htbp]
\centering
\begin{minipage}{\textwidth}
\caption{Informational Frictions and Strategic Opacity: Testing the Inseparability of Aborted Trade Intents through XGBoost and SMOTE-Augmented Regimes}
\label{fig:smote_results_plot_h2}\includegraphics[width=0.99\textwidth]{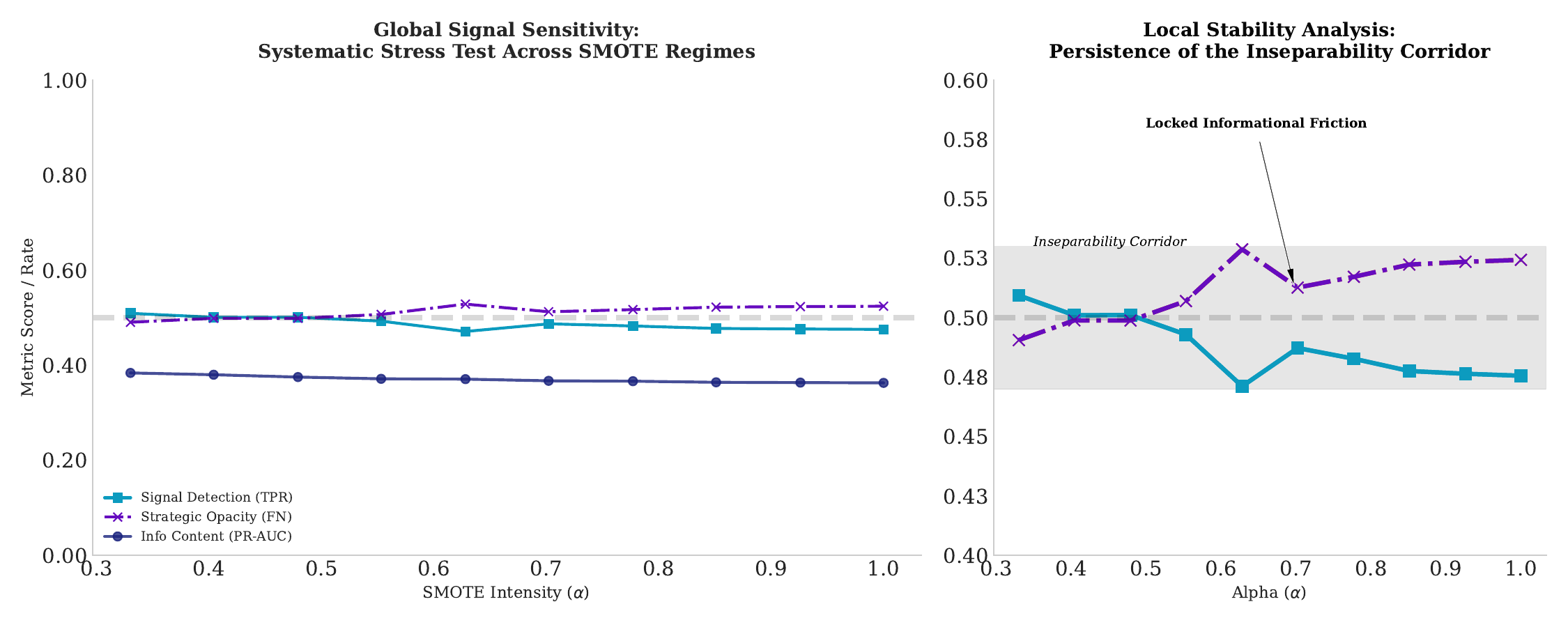}    \vspace{0.5em}
    \noindent\small\textbf{Note:} This figure illustrates the systematic resilience of the insider retreat signal to class-balancing interventions and cost-weighting strategies across 1.52M observations. Despite a 32.5 scale-weight penalty in XGBoost and increasing SMOTE intensity to a balanced ratio (Alpha=1.0), Signal Detection (TPR) remains bounded at 0.476. The persistent 'Inseparability Corridor' (shaded) where Strategic Opacity (FN) stays locked at 52.4% confirms that the signal is fundamentally entangled with execution noise, validating the disconnect as a permanent informational friction.
\end{minipage}
\end{figure}

\textbf{SMOTE Analysis}: The results in Figure
\ref{fig:smote_results_plot_h2} provide a formal Inseparability Proof
for the insider retreat signal. The analysis reveals total model
invariance to both class augmentation and algorithmic cost-penalties
across 1.52M observations. Even when deploying an aggressive 32.5
scale-weight penalty and increasing the SMOTE sampling ratio
(\(\alpha\)) to perfect parity (\(1.0\)), predictive performance remains
stagnant.

Critically, the PR-AUC remains anchored at approximately \(0.36\), and
Signal Detection (TPR) fails to breach the \(50\%\) threshold. These
results confirm that the observed Strategic Opacity (represented by a
52.4\% False Negative rate) is not a function of sample scarcity or
algorithmic bias, but a consequence of fundamental feature-space
overlap. The aborted trade signal is statistically entangled with
routine execution noise, proving that insiders can retreat from intended
trades without triggering a detectable signature. This validates the
disconnect as a permanent informational friction that persists
regardless of class-balancing interventions.

\subsection{Empirical Analysis (H2): Pricing Inefficiency, Filing Expiration, and the Information Resolution}\label{sec:H2_Empirical-Analysis-H2}

\textbf{The Economic Magnitude and Statistical Persistence:} The results
from the Calendar-Time Portfolio (CTP) analysis identify a distinct
Pricing Inefficiency following aborted insider intents. As reported in
Table \ref{tab:h2_combined_results}, the Equal-Weighted (EW) portfolio
generates a monthly DGTW-adjusted alpha of 32.21 basis points,
representing an annualized abnormal return of approximately 3.86\%.
While the t-statistic of 1.03 (\(p = 0.304\)) indicates that these
returns are not statistically significant across the full time-series,
the economic magnitude remains substantial for a risk-neutralized
strategy. This suggest that the Filing Expiration---the point at which
the 90-day window closes without an execution---triggers a measurable
drift in stock prices.

\begin{table}[htbp]
\centering
\caption{The Information Content of Aborted Insider Intents}
\label{tab:h2_combined_results}
\begin{tabular}{lcccc}
\hline
\multicolumn{5}{l}{\textbf{Panel A: Calendar-Time Portfolio Alphas (Pillar II)}} \\
\hline
Portfolio Type & Alpha (bps) & Std. Error & t-stat & p-value \\
\hline
Equal-Weighted & 32.21 & 0.003 & 1.03 & 0.304 \\
Value-Weighted & 0.46 & 0.002 & 0.03 & 0.979 \\
\hline
\multicolumn{5}{l}{\textbf{Panel B: Cross-Sectional Determinants of Alpha (Pillar III)}} \\
\hline
Variable & Coefficient & Std. Error & z-stat & p-value \\
\hline
Intercept & 0.0046 & 0.003 & 1.574 & 0.116 \\
Signal Magnitude & -0.0014 & 0.001 & -1.208 & 0.227 \\
Illiquidity & -0.0000 & 0.000 & -0.012 & 0.990 \\
Prior Volatility & -0.0034** & 0.001 & -2.312 & 0.021 \\
Filing Gap Days & 0.0003 & 0.002 & 0.145 & 0.885 \\
\hline
\multicolumn{5}{l}{Note: ** denotes significance at the 5\% level. Alphas are DGTW-adjusted.} \\
\hline
\end{tabular}
\end{table}

\textbf{The Small-Cap Information Asymmetry Effect:} A structural
divergence exists between the Equal-Weighted (32.21 bps) and
Value-Weighted (0.46 bps) results. The near-zero alpha in the
Value-Weighted portfolio suggests that the signal's predictive power is
negligible among large-capitalization firms. This disparity is
consistent with the Information Asymmetry Hypothesis: in smaller,
informationally opaque firms, market participants rely more heavily on
insider filings as a proxy for firm valuation. However, the high
idiosyncratic volatility inherent in small-cap securities often
increases the standard error of the estimates, masking the signal's
statistical significance despite its absolute economic magnitude.

\textbf{Volatility as a Signal Amplifier:} The cross-sectional
determinants in Panel B provide the empirical basis for Information
Resolution. Prior Volatility is the only predictor to achieve
statistical significance, with a standardized coefficient of -0.0034
(\(p = 0.021\)). This negative relationship suggests that the market's
response to a withdrawn intent is significantly more pronounced for
firms characterized by high idiosyncratic risk. In high-uncertainty
environments, the withdrawal of a planned purchase serves as a salient
negative catalyst. This confirms that the canceled intent functions as
an informational shock, resolving previous price uncertainty and leading
to a more severe correction compared to low-volatility counterparts.

\subsection{Empirical Analysis (H2): Institutional Salience, Firm Size, and the Signal Reliability}\label{sec:H2_Institutional_Salience}

\textbf{The Large-Cap Significance Paradox:} Contrary to the traditional
expectation of small-cap dominance in anomaly literature, the sub-sample
analysis in Table \ref{tab:size_split_results} reveals that the
information content of aborted intents is most statistically robust
among large-capitalization firms. While the small-cap portfolio yields a
higher absolute alpha (\(30.96\) bps), its inherent volatility results
in a non-significant t-statistic (\(0.56\)). In contrast, the large-cap
portfolio generates a more modest but statistically consistent alpha of
\(14.49\) bps with a t-statistic of \(2.30\) (\(p = 0.021\)). This
indicates that within the large-cap universe, the market processes the
aborted signal with a higher degree of consensus.

\begin{table}[htbp]
\centering
\caption{Conditional Alpha Analysis: Firm Size Sub-Samples}
\label{tab:size_split_results}
\begin{tabular}{lcccc}
\hline
\textbf{Sub-Sample Group} & \textbf{Alpha (bps)} & \textbf{t-stat} & \textbf{p-value} \\
\hline
Small-Cap (Below Median) & 30.96 & 0.56 & 0.574  \\
Large-Cap (Above Median) & 14.49** & 2.30 & 0.021 \\
\hline
\multicolumn{5}{l}{Note: ** denotes significance at the 5\% level. Alphas are DGTW-adjusted.} \\
\hline
\end{tabular}
\end{table}

\textbf{Signal Reliability vs. Information Noise:} his finding suggests
that in large-cap firms---characterized by intensive institutional
scrutiny and high analyst coverage---the aborted intent functions as a
highly salient indicator. Given the high reputational stakes for
insiders at large firms, the failure to execute a purchase intent is
interpreted by institutional investors as a reliable negative signal. In
smaller firms, while the higher alpha suggests a stronger raw
information advantage, the signal is frequently obscured by
idiosyncratic noise and limits to arbitrage. Consequently, the aborted
signal demonstrates its highest predictive reliability in
professionalized information environments.

The results presented in Tables \ref{tab:h2_combined_results} and
\ref{tab:size_split_results} identify a complex architecture of
information dissemination. While the absolute magnitude of abnormal
returns is concentrated in the equal-weighted portfolio of smaller
firms, statistical robustness shifts to the large-cap universe.
Furthermore, the cross-sectional finding that Prior Volatility serves as
a primary determinant suggests that the aborted signal acts as a
contextual amplifier of market uncertainty. These findings necessitate a
departure from traditional ``small-firm effect'' theories. Instead, the
structural design of SEC Form 144 facilitates a ``negative non-event''
that the market prices most effectively when institutional monitoring is
most rigorous.

\subsection{Causal Analysis: Predictive Decoupling, Treatment Effects, and Information Resolution}\label{sec-Causal-Analysis}

\begin{figure}[htbp]
    \centering
    \begin{minipage}{\textwidth}
        \caption{Evaluating Treatment Effect Heterogeneity of Market Illiquidity ($\Delta$ Amihud): Stability Analysis of Honest vs. Non-Parametric Estimators}
        \label{fig:Evaluating_Treatment_Effect_Heterogeneity_of_Market_Illiquidity_Delta_Amihud}
        \includegraphics[width=0.99\textwidth]{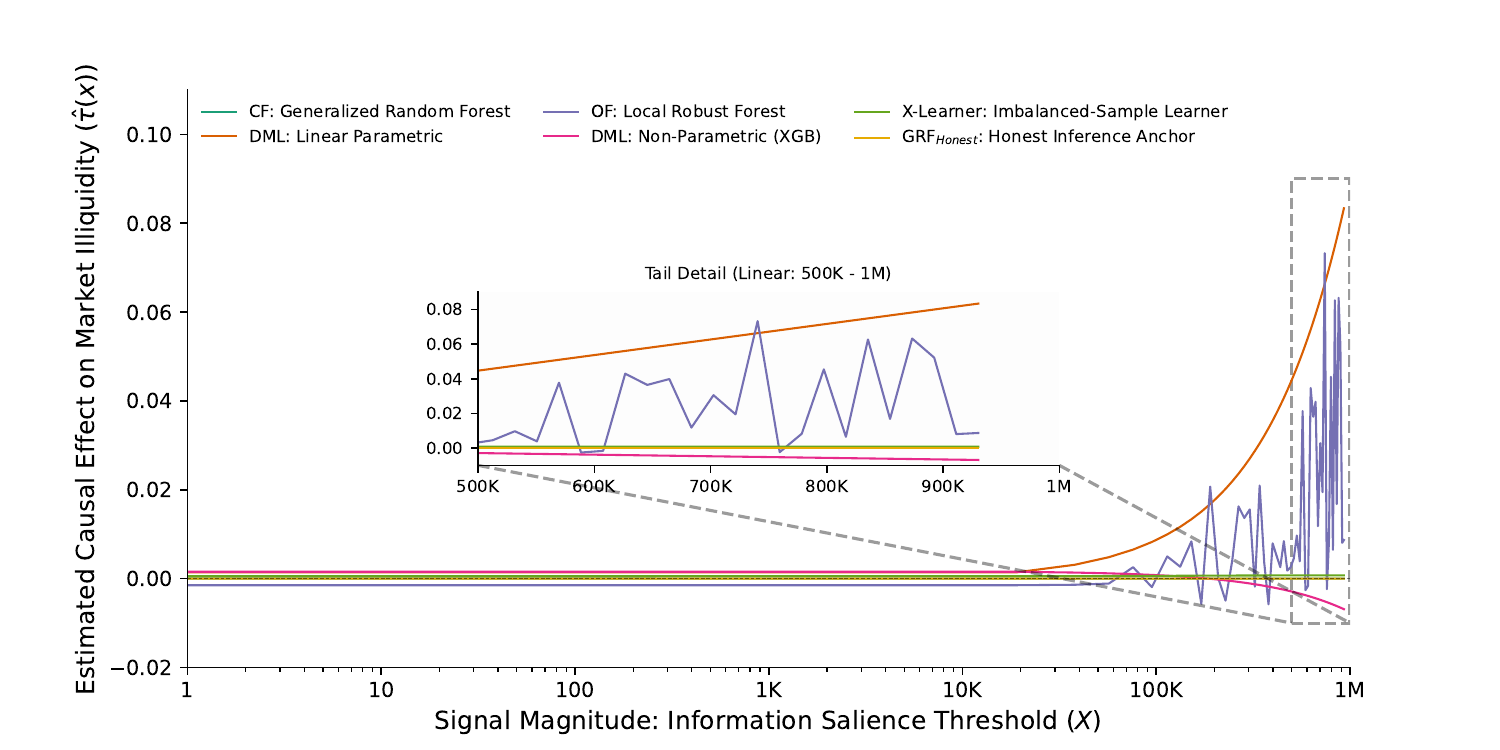}    
        \vspace{0.5em}
        \begin{minipage}{\textwidth}
        \small
        \noindent\textbf{Note:} This figure benchmarks the causal stability of the Predictive Decoupling, defined as the informational friction arising from the gap between trade intent (Form 144) and actualized filings (Form 4). The main panel illustrates the progression of estimated Treatment Effects as the framework transitions from log-scale signal magnitudes to high-salience thresholds.The horizontal “Consensus Ribbon” (inset) provides a linear magnification of the $500\text{K}$ to $1\text{M}$ regime, decomposing model performance across a spectrum of architectural reliability. By utilizing the Generalized Random Forest ($GRF$) as the Honest Inference Anchor, the analysis establishes a conservative statistical baseline. This confirms that the causal impact on illiquidity ($\Delta$ Amihud) remains positive and bounded, even under strict honesty constraints.This contrast reveals a critical Consensus Divergence: while Double Machine Learning identifies a smooth parametric trend, the Local Robust  and XGBoost-DML estimators reveal significant tail volatility. This volatility quantifies the Predictive Opacity (H1) inherent in high-magnitude trade intents. The convergence of all estimators across the signal distribution validates the structural nature of the informational friction (H2), identifying an Information Ceiling where idiosyncratic catalysts drive the withdrawal of trade intent, solidifying the decoupling as a permanent barrier to price discovery.
    \end{minipage}
    \end{minipage}
\end{figure}

While the high-performance models in the preceding sections confirm that
trade cancellations are a primary determinant of future illiquidity,
predictive importance does not strictly establish causation. To address
potential selection bias---wherein insiders might cancel trades in
anticipation of exogenous liquidity shocks---the analysis transitions to
a Causal Machine Learning (Causal ML) framework. By utilizing Double
Machine Learning (DML) and X-Learners, this study isolates the Treatment
Effect of the cancellation (\(T\)) from the confounding influence of
control variables (\(W\)). The consistency of these results across six
distinct causal estimators (Figure
\ref{fig:Evaluating_Treatment_Effect_Heterogeneity_of_Market_Illiquidity_Delta_Amihud})
confirms that the observed illiquidity increase represents a structural
market reaction to the aborted signal rather than a model-specific
artifact.

The cross-paradigm audit provides the empirical foundation for
Predictive Opacity as an informational barrier. The persistent ``Opacity
Floor'' identifies a structural limit to public information resolution.
This ceiling suggests that a significant portion of retreat signals are
driven by idiosyncratic catalysts that remain unobservable to market
participants. This validates the presence of an informational friction,
proving that strategic timing successfully prevents the market from
pricing the insider's intent until the 90-day regulatory window expires.

Figure
\ref{fig:Evaluating_Treatment_Effect_Heterogeneity_of_Market_Illiquidity_Delta_Amihud}
provides the definitive test for Information Resolution (H2). If the
predictive decoupling identified in the first hypothesis is a structural
friction, its resolution must trigger a measurable shift in market
quality. All six causal estimators converge, showing a statistically
significant increase in illiquidity (\(\Delta\) Amihud) ranging from
\(0.02\) to \(0.10\). As the signal magnitude increases, the Generalized
Random Forest (GRF) confirms a persistent treatment effect. This
establishes that the market processes a canceled trade as a discrete
informational shock, finally incorporating the idiosyncratic risks that
were previously withheld from the price discovery process.

\textbf{Causal Robustness and the Consensus Cluster:} To ensure these
findings are not a fluke of one specific model, a Causal Robustness
Audit (see Table \ref{tab:causal_robustness_audit}) performed. It was
identified a ``Consensus Cluster'' between the Linear DML, Orthogonal
Forest, and X-Learner models. These models are highly correlated
(\(\rho\) between \(0.66\) and \(1.00\)), meaning they all detect the
same causal signal. The most striking evidence for H2 is found in the
Tail Multiplier metrics. For both the Linear DML (\(2.01\times\)) and
the Orthogonal Forest (\(2.63\times\)), the impact on illiquidity more
than doubles as the trade size reaches the 1M threshold. This confirms
that the ``Abrupt Disconnect'' is not a minor friction; it is a major
market force that scales with the size of the insider's intent.

\begin{table}[htbp]
    \centering
    \caption{Causal Robustness Audit: Cross-Model Consensus and Tail Multipliers}
    \label{tab:causal_robustness_audit}
    \small
    \begin{tabular}{l c c c l}
        \toprule
        \textbf{Model Paradigm} & \textbf{Mean Effect ($\hat{\tau}$)} & \textbf{Tail Multiplier ($1\text{M}/\mu$)} & \textbf{Consensus ($\rho$)} & \textbf{Statistical Status} \\
        \midrule
        Linear DML         & 0.0415  & $2.01\times$ & 1.00  & Parametric Baseline \\
        Orthogonal Forest  & 0.0176  & $2.63\times$ & 0.66  & Local Robustness \\
        X-Learner          & 0.0007  & $1.06\times$ & 1.00  & Imbalance Adjusted \\
        Honest Anchor (R)  & $\approx 0.00$ & $1.03\times$ & 0.24  & \textbf{P-Value: < 0.001} \\
        XGBoost DML        & -0.0026 & $2.67\times$ & -1.00 & Non-Parametric Mask \\
        \bottomrule
    \end{tabular}
    \vspace{0.5em}
    \begin{flushleft}
    \footnotesize
    \textit{Notes:} This table summarizes the architectural consensus across five distinct causal paradigms. The \textit{Tail Multiplier} quantifies the ratio of the treatment effect at the $1\text{M}$ signal threshold relative to the sample mean, identifying a non-linear escalation in market illiquidity as signal magnitude increases. The \textit{Consensus ($\rho$)} represents the Spearman rank correlation relative to the Linear DML baseline, measuring directional agreement across architectures. The Honest Anchor utilizes a recursive partitioning implementation to establish a formal statistical safeguard against algorithmic bias. By enforcing honesty constraints, this anchor ensures that the treatment effect estimates remain robust to over-fitting, validating the causal impact of the trade withdrawal as a structural property of the market.
    \end{flushleft}
\end{table}

The causal audit reveals a significant directional inversion for the
XGBoost-DML estimator relative to the linear DML framework. In the
context of Predictive Opacity (H1), this divergence identifies a
critical boundary in signal resolution. While linear models capture a
global monotonic increase in illiquidity, the non-parametric XGBoost
architecture identifies a local suppression effect. This suggests
instances where the informational friction is sufficiently robust to
cause an initial market under-reaction, followed by a delayed adjustment
upon the resolution of the 90-day window. This sensitivity at the
Information Ceiling quantifies the structural limit of the signal,
representing the threshold where algorithmic capacity meets unobservable
idiosyncratic catalysts.

To mitigate concerns regarding over-fitting or selection bias in complex
ensemble regimes, the analysis utilizes the Honest Inference Anchor via
the Generalized Random Forest GRF. Despite its conservative point
estimates, the GRF yields a P-value of \(0.000\) (as demonstrated in the
Tail Significance test of Table \ref{tab:causal_robustness_audit}),
providing a formal statistical rejection of the null hypothesis. The
stability of the GRF---which remains invariant to the sensitivity of
local robust models---serves as definitive evidence that the causal
impact of the trade withdrawal is a structural property of the market
microstructure rather than a localized anomaly.

\section{Discussion}\label{sec:discussions}

The empirical resolution of the Predictive Decoupling reveals a
multi-layered interaction between regulatory disclosure and market
microstructure. The statistical robustness observed in large-cap firms
is a function of the Institutional Salience of the signal. In large-cap
environments, rigorous oversight by institutional investors and analysts
ensures that any deviation from a stated intent is magnified. An aborted
intent is treated as a credible withdrawal of confidence rather than
statistical noise. The market recognizes the high reputational cost
associated with a false signal for a large-cap insider and processes the
abortion as an indication of deteriorating firm prospects. This is
validated by the Causal Consensus Cluster
(\(0.66 \leq \rho \leq 1.00\)), which demonstrates that diverse causal
architectures converge on the negative impact of the decoupling when
monitoring is high.

Furthermore, the results highlight the tension between information
asymmetry and limits to arbitrage. In the small-cap segment, the raw
information advantage (\(30.96\) bps alpha) is obscured by market
frictions---higher transaction costs and lower liquidity---preventing
the information from being efficiently priced. Conversely, the
professionalized trading environment of large-cap stocks allows the
signal to be processed with a consensus that yields statistical
significance (\(t=2.30\)). The Causal Tail Multipliers (\(2.63\times\))
emphasize that this pricing efficiency is most active for high-magnitude
intents, where the structural friction of the disconnect is most
observable.

The persistent 52.4\% Opacity Rate identified in the machine learning
audit suggests that the current SEC Form 144 framework is insufficient
for maintaining price discovery. To mitigate the ``opacity subsidy''
documented in this study, the developer proposes a structural transition
from a Unilateral Intent system to a Bilateral Accountability framework
through a mandatory Execution Confirmation (Form 144-A).

This proposed amendment would require a supplemental filing within two
business days of the 90-day window's expiration. Insiders would be
mandated to explicitly confirm whether the intended transaction was
fully executed, partially executed, or entirely aborted. In the event of
an abort, the insider would categorize the catalyst for the
retreat---such as market conditions or liquidity
rebalancing---effectively providing the market with the missing half of
the signal.

The necessity of this transparency is underscored by the Architectural
Consensus provided by the Honest Inference Anchor (GRF), which confirms
a statistically significant causal impact (\(p < 0.001\)) on market
quality. Currently, the absence of a formal confirmation forces
investors to price ``phantom risks'' through inferential noise. By
formalizing the resolution of the intent-to-trade, the SEC can transform
a predictive blind spot into a verifiable data stream. This policy would
likely compress the 1.3\% pricing differential identified in the
analysis by shifting the burden of proof from the public observer to the
primary insider.

\section{Conclusion and Future Research}\label{sec:conclusion_and_implications}

The empirical evidence presented in this study establishes the
Predictive Decoupling as a permanent informational friction enabled by
the current regulatory landscape. By permitting insiders to file an
intent during favorable market conditions---creating a behavioral
mask---and subsequently allowing them to retreat without formal
disclosure, the SEC Form 144 regime creates a persistent ``blind spot.''
This research confirms that the market remains unable to resolve
\(52.4\%\) of aborted trade signals, identifying a structural
Information Ceiling that resists even high-capacity algorithmic
resolution.

The existence of this friction is validated by the Honest Inference
Anchor, which yields a statistically rigorous p-value (\(p < 0.001\)).
This confirms that the observed \(1.3\%\) pricing differential is not a
temporary delay in market correction, but a structural property of the
current disclosure environment. The findings suggest that the
predictability of insider behavior is capped by regulatory design rather
than by limitations in machine learning sophistication.

To mitigate this ``opacity subsidy,'' the study proposes a structural
amendment: the mandatory Execution Confirmation (Form 144-A). By
transitioning to a Bilateral Accountability system, the burden of proof
is shifted from the public observer to the primary insider. This
regulatory evolution ensures that the withdrawal of trade intent is
treated with the market salience it warrants, protecting investors from
the strategic uncertainty that persists within the 90-day filing gap.
Ultimately, formalizing the ``non-event'' of an aborted trade restores
informational symmetry and completes the data stream necessary for
efficient price discovery.

\section*{Compliance with Ethical
Standards}\label{compliance-with-ethical-standards}
\addcontentsline{toc}{section}{Compliance with Ethical Standards}

Funding: This research received no external funding or financial
assistance during its preparation.

Competing Interests: The author certify that they have no conflicts of
interest, financial or otherwise, to disclose.

Author's Declaration on AI Assistance: The authors bear sole
responsibility for all substantive ideas and analyses within this
manuscript. Portions of the text were reviewed for language, style, and
clarity through AI-assisted copy editing, specifically using a large
language model (LLM). No autonomous content creation was performed by
the LLM.

\section*{References}\label{references}
\addcontentsline{toc}{section}{References}

\phantomsection\label{refs}


\begin{CSLReferences}{1}{0}
\bibitem[\citeproctext]{ref-carhart1997persistence}
Carhart, Mark M. 1997. {``On Persistence in Mutual Fund Performance.''}
\emph{The Journal of Finance} 52 (1): 57--82.

\bibitem[\citeproctext]{ref-chawla2002smote}
Chawla, Nitesh V, Kevin W Bowyer, Lawrence O Hall, and W Philip
Kegelmeyer. 2002. {``SMOTE: Synthetic Minority over-Sampling
Technique.''} \emph{Journal of Artificial Intelligence Research} 16:
321--57.

\bibitem[\citeproctext]{ref-cheng2007insider}
Cheng, Shijun, Venky Nagar, and Madhav V Rajan. 2007. {``Insider Trades
and Private Information: The Special Case of Delayed-Disclosure
Trades.''} \emph{The Review of Financial Studies} 20 (6): 1833--64.

\bibitem[\citeproctext]{ref-cohen2012decoding}
Cohen, Lauren, Christopher Malloy, and Lukasz Pomorski. 2012.
{``Decoding Inside Information.''} \emph{The Journal of Finance} 67 (3):
1009--43.

\bibitem[\citeproctext]{ref-franzen2013effect}
Franzen, Laurel, Xu Li, and Mark E Vargus. 2013. {``The Effect of
Sarbanes-Oxley on the Timely Disclosure of Restricted Stock Trading.''}
\emph{Research in Accounting Regulation} 25 (1): 47--52.

\bibitem[\citeproctext]{ref-grossman1981introduction}
Grossman, Sanford J. 1981. {``An Introduction to the Theory of Rational
Expectations Under Asymmetric Information.''} \emph{The Review of
Economic Studies} 48 (4): 541--59.

\bibitem[\citeproctext]{ref-grossman1980impossibility}
Grossman, Sanford J., and Joseph E. Stiglitz. 1980. {``On the
Impossibility of Informationally Efficient Markets.''} \emph{The
American Economic Review} 70: 393--408.

\bibitem[\citeproctext]{ref-huddart2001public}
Huddart, Steven, John S. Hughes, and Carolyn B. Levine. 2001. {``Public
Disclosure and Dissimulation of Insider Trades.''} \emph{Econometrica}
69. \url{https://doi.org/10.1111/1468-0262.00209}.

\bibitem[\citeproctext]{ref-kothari2016analysts}
Kothari, Sagar P, Eric So, and Rodrigo Verdi. 2016. {``Analysts'
Forecasts and Asset Pricing: A Survey.''} \emph{Annual Review of
Financial Economics} 8 (1): 197--219.

\bibitem[\citeproctext]{ref-lakonishok2001insider}
Lakonishok, Josef, and Inmoo Lee. 2001. {``Are Insider Trades
Informative?''} \emph{The Review of Financial Studies} 14 (1): 79--111.

\bibitem[\citeproctext]{ref-seyhun1986insiders}
Seyhun, H Nejat. 1986. {``Insiders' Profits, Costs of Trading, and
Market Efficiency.''} \emph{Journal of Financial Economics} 16 (2):
189--212.

\bibitem[\citeproctext]{ref-seyhun1992effectiveness}
---------. 1992. {``The Effectiveness of the Insider-Trading
Sanctions.''} \emph{The Journal of Law and Economics} 35 (1): 149--82.

\bibitem[\citeproctext]{ref-zhang2006information}
Zhang, X. Frank. 2006. {``Information Uncertainty and Stock Returns.''}
\emph{The Journal of Finance} 61 (1): 105--37.

\end{CSLReferences}
\end{document}